\documentclass[pdftex,twocolumn,epjc3]{svjour3}          

\RequirePackage[T1]{fontenc}

\smartqed  

\RequirePackage{graphicx}
\RequirePackage{amsmath,amssymb}
\RequirePackage{newtxmath,newtxtext}
\RequirePackage{flushend}
\RequirePackage[numbers,sort&compress]{natbib}
\RequirePackage{soul,xcolor}
\sethlcolor{yellow}

\RequirePackage[colorlinks,citecolor=blue,urlcolor=blue,linkcolor=blue]{hyperref}

\journalname{Eur. Phys. J. C}

\begin{document}
	\sloppy

\title{Holographic dark energy from the laws of thermodynamics with Rényi entropy}

\author{Manosh T. Manoharan\thanksref{e1,addr1}
        \and
        N. Shaji\thanksref{e2,addr1}
        \and
        Titus K. Mathew\thanksref{e3,addr1,addr2}}

\thankstext{e1}{tm.manosh@gmail.com, tm.manosh@cusat.ac.in}
\thankstext{e2}{shajin@cusat.ac.in}
\thankstext{e3}{titus@cusat.ac.in}

\institute{Department of Physics, Cochin University of Science and Technology, Kochi, India -- 682022\label{addr1}
          \and
         Centre for Particle Physics, Cochin University of Science and Technology, Kochi, India -- 682022\label{addr2}
}

\date{Received: date / Accepted: date}

\maketitle

\begin{abstract}
This article investigates the relationship between the holographic principle and the laws of thermodynamics in explaining the late-time acceleration of the universe. First, we explore the possibilities of generating the standard holographic dark energy (SHDE) from the laws of horizon thermodynamics. Except for entropies that follow an exponent stretched area law, unless we redefine the horizon temperature, we found it challenging to construct a one-to-one correspondence between the dark energies defined by the holographic principle and the laws of thermodynamics. Secondly, in SHDE models, unless we invoke some phenomenological interactions, it is impossible to explain the late-time cosmic acceleration with the Hubble horizon as the IR cutoff. On the other hand, it is possible to induce dark energy as an integration constant using the laws of thermodynamics on the Hubble horizon. These motivated us to explore a feasible way to invoke the holographic principle from the laws of horizon thermodynamics. We show that the additional terms that appear in the modified Friedmann equations on using entropies other than the Bekenstein-Hawking entropy in the first law of thermodynamics can behave like a dynamic holographic dark energy (HDE). We study the features of such an HDE with R\'enyi entropy as the choice without considering any non-standard interactions. Interestingly, the resulting form of dark energy reduces to the standard cosmological constant when R\'enyi entropy reduces to the Bekenstein-Hawking entropy. By examining different parameters, we affirm the validity of our approach to dark energy, which respects both holographic principle and thermodynamics. 
\end{abstract}

\keywords{Dark energy, Holographic principle, Laws of thermodynamics, R\'enyi entropy}

\section{\label{sec:Introduction}Introduction}

Explaining the recent accelerated expansion of the universe is one of the significant theoretical endeavours in modern cosmology \cite{Perlmutter_1997, Riess_1998}. A possible explanation requires a cosmic component called the ``\textit{dark energy}'' with negative pressure, whose true nature is still obscure. The simplest candidate for dark energy is the cosmological constant, advocated by the $\Lambda$CDM model, which should be same as the energy density of the vacuum \cite{Condon_2018, RevModPhys.75.559}. However, the vacuum energy density predicted by the standard quantum field theory (QFT) and the observed dark energy density have a discrepancy of $\sim120$ orders in magnitude \cite{RevModPhys.61.1}. This mismatch leads to the fine-tuning problem. Further, there is no conclusive explanation for why the present value of dark energy density has the same order of magnitude as that of the matter component, other than being a coincidence \cite{PhysRevLett.82.896}. A practical solution to these problems is to replace the cosmological constant with a dynamical dark energy \cite{doi:10.1142/S021827180600942X, SHAPIRO2009105}. There are many efforts to explain the observed data with several dynamical dark energy models \cite{10.1093/mnras/staa3914, PhysRevD.104.023510}, and more conclusions will follow as we improve our observations~\cite{aghamousa2016desi}.

Among numerous dynamical dark energy models, a promising approach comes from the holographic principle. In its most general sense, the holographic principle refers to the duality between theories of the bulk and its boundary \cite{hooft2009dimensional, doi:10.1063/1.531249, doi:10.1142/5689,  Maldacena1999, RevModPhys.74.825}. This notion of dimensional reduction proposed by 't Hooft \cite{hooft2009dimensional}, based on the non-extensive scaling of black hole entropy, prevents one from over-counting the degrees of freedom. The disparity mentioned above between the measured and the predicted values of the dark energy density is because local quantum field theory over-counts the degrees of freedom. Accounting the holographic principle, Cohen, Kaplan and Nelson \cite{PhysRevLett.82.4971} conjectured that an effective field theory which connects the energy density and length scale through a saturation entropy could adequately describe the observed dark energy density (${\rho_{_\Lambda}}$). More precisely, their conjecture reads,
\begin{equation}
{\rho_{_\Lambda}}\leq\frac{S}{L^4}.
\label{eq:CKNBound}
\end{equation}
Here, $S$ is the entropy of the cosmological horizon, proportional to its surface area as motivated by black hole physics \cite{PhysRevD.7.2333, PhysRevD.9.3292,PhysRevD.13.191, Hawking1975, HAWKING1974, PhysRevD.15.2738, PhysRevD.14.870, Bardeen1973}, and $L$ is the length scale called the infra-red (IR) cutoff. The IR cutoff is generally a function of the Hubble parameter and its time derivatives \cite{GRANDA2008275}. Thus, the non-extensive entropy associated with the horizon connects to the entropy of the ``dark energy'' by the notion of holography. Many authors refer to this relation as the Cohen-Kaplan-Nelson (CKN) bound or simply the ``\textit{holographic principle}'' \cite{PhysRevLett.82.4971, PhysRevD.101.126010,PhysRevD.104.076024}. Although, the term ``\textit{holographic principle}'' is widely used in various areas, here we refer to the CKN relation given above.

In standard holographic dark energy (HDE) models, one identifies the dark energy density as the vacuum energy density of the underlying effective field theory and proposes a dynamical expression from the above CKN bound as,
\begin{equation}
\rho_{_{\Lambda }}\propto\frac{S}{L^4}.
\label{eq:CKNHDE_original}
\end{equation}\textit{}
The advantage of the HDE model is that it favours a very small value for the current vacuum energy density \cite{PhysRevLett.85.1610} and is also stable against divergent quantum corrections \cite{PhysRevLett.89.081301}. Lately, Banks and Draper \cite{PhysRevD.101.126010} suggested modifications to the CKN relation in cosmology based on the Nariai limit. Originally, the CKN bound was an attempt to establish a correlation between the IR and UV cutoffs in quantum field theories. However, there are new proposals which give different interpretation to the same. For example, Blinov and Draper gives an interpretation in which the QFT degrees of freedom are depleted as a function of scale \cite{PhysRevD.104.076024}. Their interpretation is consistent with the apparent success of standard QFT in particle physics and its failure at the cosmic scale. It is important to note that Cohen et al. \cite{PhysRevLett.82.4971} proposed the CKN bound for a $3+1$ dimensional spacetime with a horizon entropy following the Bekenstein-Hawking area law \cite{HAWKING1974,PhysRevD.7.2333,Hawking1975}. Reference \cite{sym13020237} proposes a generalization of this bound to discrete higher dimensions. Many authors investigated the implications of this holographic principle in cosmology \cite{ PhysRevD.105.034022, PhysRevD.103.026005, PhysRevD.102.123537,  PhysRevD.99.026002, PhysRevD.93.026006, Miao2015, PhysRevD.104.123519, sym13060928, Nojiri2006, PhysRevD.102.023540} and other areas of physics \cite{davoudiasl2021cohen, Draper2022SnowmassWP}. The above references give solid arguments to adapt the holographic principle and the corresponding dynamical dark energy $\left(\rho_{\Lambda}\propto S/L^4\right)$ in cosmology. For instance, in \cite{PhysRevD.102.123525}, Saridakis demonstrated the validity of the principle with Barrow entropy as the choice.

The connection between horizon thermodynamics and holography \cite{Wang2018} led Moradpour et al. \cite{Moradpour2018} to reason that a proper connection between ${\rho_{_\Lambda}}$, $L$ and $S$ should also come from the laws of thermodynamics. Recently, reference \cite{Ramakrishna_2021} pointed out a thermodynamic origin for the CKN bound based on the principle of free energy minimization. To propose the thermodynamic connection, Moradpour et al. assumed that the total energy ($E_T$) during the late phase of the universe mostly consists of dark energy ($E_{_\Lambda}$) alone. Then, using the relation $E_{_\Lambda}\sim\rho_{_{\Lambda}}V\simeq E_T\propto TS$, where $T$ is the horizon temperature and $V$ is the areal volume bounded by the horizon, they proposed,
\begin{equation}
{\rho_{_\Lambda}}\propto T\frac{dS}{dV},
\label{eq:FLTBound}
\end{equation}
as the expression for dark energy density. The immediate question is whether there is a one-to-one correspondence between this expression and the HDE density from the CKN bound. If yes, then equation (\ref{eq:FLTBound}) and (\ref{eq:CKNHDE_original}) must be at least proportional to each other or there should exist some function relation between them. In this article, we show that one cannot confirm such a proportionality or some functional correspondence for all choices of entropy with $\rho_{_{\Lambda}}$ coming from the first law of thermodynamics.

Prior to Moradpour et al.'s proposal, Luongo proposed a thermodynamic approach to holographic dark energy \cite{luongo2017thermodynamic}.  They related the holographic minimal information density to the de Broglie wavelength at a given temperature. They demonstrate the existence of an energy density that appears in the energy-momentum tensor, which they interpret differently from the vacuum energy. The essence of their approach lies in finding a proper thermal cutoff associated with massive and massless particles. Although Luongo's approach is based on the holographic principle and laws of thermodynamics, it is different from what we follow in this manuscript, which is based on the conjecture proposed by Cohen et al. \cite{PhysRevLett.82.4971}.

In proposing the relation (\ref{eq:FLTBound}), the authors in \cite{Moradpour2018} assumed a flat $3+1$ dimensional FLRW universe and the validity of the standard Friedmann equations. However, Golanbari~et al. \cite{golanbari2020renyi} refuted the cosmological model proposed in \cite{Moradpour2018} and pointed out that one must use the modified Friedmann equations when using any generalised entropies. Although reference~\cite{golanbari2020renyi} considered a modified version of Friedmann equations to study cosmic evolution, they used the same definition for dark energy proposed in \cite{Moradpour2018}, which according to our calculations, already assumes the validity of the standard Friedmann equations. Together, references \cite{golanbari2020renyi} and \cite{Moradpour2018} illustrates the inconsistency between the conventional holographic and the thermodynamic approaches in cosmology. This disparity led us to reconsider the following questions.

\begin{quote}
	``\emph{Is it possible to model dark energy density from the laws of thermodynamics, which is consistent with the holographic principle? If yes, how? and if no, why not?}''
\end{quote}
Our answer is ``yes'', and the rest of this manuscript answers ``how''.

Further, there is no argument that the $\Lambda$CDM model is the simplest model that can very well explain the observational data, and so far, no dynamical dark energy models have shown strong evidence against it. Then, it is only reasonable to assume that, if at all dark energy is dynamic, any new approach must closely resemble the $\Lambda$CDM model. We show that such a model is possible by combining the notion of holography with the laws of thermodynamics. Additionally, one can see that Moradpour et al.'s proposal is very similar to the entropic force formalism by Verlinde \cite{Verlinde2011}. It is thus essential to investigate the possible correlations between entropic dark energy, $\Lambda$CDM and HDE models from the perspective of thermodynamics \cite{PhysRevD.86.043010}. Considering all these statements, we found that the additional entropic functions that appear in the modified Friedmann equations can take up the role of holographic dark energy (HDE) to explain cosmic evolution. We study the features of such an HDE with R\'enyi entropy as the choice without considering any non-standard interactions. Interestingly, our form of dark energy reduces to the standard cosmological constant when R\'enyi entropy reduces to the Bekenstein-Hawking entropy. By examining different parameters, we affirm the validity of our approach to dark energy, which respects holography and thermodynamics.

The structure of the article is as follows. First we review the notion of standard HDE and its thermodynamic analogue proposed by Moradpour et al.. We then compare and contrast between these two notions using different choices of entropies. Since we found difference between the SHDE and Moradpour et al.'s proposal, we investigated the underlying reasons for the differences. Further, we explored the possible expressions for dark energy from the Clausius relation and the unified first law. We found that the new expression is also different from Moradpour et al.'s proposal and the CKN bound. The calculation confirms that the thermodynamic picture of dark energy density is not the same as the conventional HDE density. We then illustrate specific issues when we blindly assume the standard HDE model with Hubble horizon as IR cutoff. To resolve these inconsistencies, we then propose a new approach to extract the dark energy component from the laws of thermodynamics, which is consistent with the holographic bound. Further, we demonstrated the validity of our approach using R\'enyi entropy, and investigated the late time cosmological behaviour. We also studied the entropy evolution and found that the new model respects the second law of thermodynamics. Finally we summarise our findings.

\section{Standard holographic dark energy and the thermodynamic analogue}

There is no conclusive evidence that quantum fields fluctuate independently over infinitely large cosmological scales \cite{PhysRevD.104.076024}. Hence, instead of a local QFT, Cohen, Kaplan and Nelson \cite{PhysRevLett.82.4971} conjectured that an effective QFT with a UV-IR connection bounded by the horizon entropy is a feasible description. In holographic dark energy models, one identifies the dynamical dark energy from the CKN bound as,
\begin{equation}
{\rho_{_\Lambda}}=\mathcal{C}\frac{S}{L^4}.
\label{eq:CKN_HDE}
\end{equation}
Here $\mathcal{C}$ is a constant, $S$ is the horizon entropy, and $L$ is the IR cutoff (all in natural units). ``\textit{The sole purpose of this approach is to replace the vacuum energy density predicted by the standard QFT with an appropriate one}''. Here, the dynamical dark energy given in eq.~(\ref{eq:CKN_HDE}) simultaneously resolves the fine-tuning and coincidence problems. In the standard HDE approach, although ${\rho_{_\Lambda}}$ depends on the choice of entropy, the Friedmann equations remain the same as in the $3+1$ Einstein's gravity. In other words, HDE modifies the stress tensor ($T_{\mu\nu}$) rather than the Einstein tensor ($G_{\mu\nu}$). 

At this juncture, we must remember that the CKN relation is a \emph{bound} that predicts a consistent value of the cosmological constant, rather than a dynamical function. The \emph{holographic principle} thus only demands the dark energy density to be $\lesssim S/L^4$, not necessarily equal or proportional to it. Further, the notion of dark energy density to vary as a function of $\sim L^{-2}$ also comes from the renormalization group flow expression given as \cite{Shapiro_2007},
\begin{equation}
\rho_{_{\Lambda }}(H)=\frac{3}{8\pi}\left[c_0+c_1 H^2+\mathcal{O}\left(H^4\right)+\cdots\right].
\end{equation}
Here, $c_0$ is a constant of integration, and the first order term is similar to the HDE with Hubble horizon as IR cutoff. If this running vacuum approach respects the holographic principle, the motivation for HDE must be to identify a term that scales like $\sim L^{-2}$ up to the first order, not necessarily equal or proportional to $L^{-2}$. 

Now, why should one approach the HDE models from the laws of thermodynamics? Since the discovery of black hole thermodynamics, many seminal works illustrated the deeper connection between the equations of gravity and thermodynamics. Works by Jacobson \cite{PhysRevLett.75.1260}, Cai and Kim \cite{Cai_2005}, Padmanabhan \cite{Padmanabhan_2010}, Akbar and Cai \cite{PhysRevD.75.084003}, Hayward et al. \cite{Hayward_1998, HAYWARD1999347} are some notable ones. Along this line, one can study the cosmic evolution using the laws of thermodynamics and extend it to Gauss-Bonnet and more general Lovelock theories of gravity~\cite{PhysRevD.75.084003, Cai_2005}. Since horizon entropy connects holography and thermodynamics, it is natural to anticipate that any quantity described in one framework should also have a description in the other. 

Considering the thermodynamic properties of the cosmological horizon \cite{Wang2018,PhysRevLett.75.1260, PhysRevD.75.084003}, Moradpour~et al. \cite{Moradpour2018} proposed a thermodynamic analogue of the standard HDE density. Assuming dark energy as the dominant component in the late epoch of cosmic evolution, they argued that the relation $E_{_T} \simeq E_{_\Lambda} \sim \rho_{_\Lambda} V$ is a good approximation, with $E_{_T}$ as the total energy, $E_{_\Lambda}$ as the dark energy, $\rho_{_{\Lambda}}$ as the dark energy density and $V$ as the areal volume. Then, they proposed a dark energy density which takes the form,
\begin{equation}
\rho_{_{\Lambda}}=\mathcal{C}_{_{\tiny T}}T\frac{dS}{dV}.
\label{eq:FLT_DE}
\end{equation}
Here $\mathcal{C}_{_{\tiny T}}$ is the proportionality constant. Similar to the standard HDE approach, the authors assumed the validity of standard Friedmann equations and only accounted for modifications to the vacuum energy density. 

Before going into the details, or verifying the validity of Moradpour et al.'s proposal, let us contrast it with the standard HDE approach. To compare, we will use different non-extensive entropies for $S$ and use the apparent horizon radius, ${\tilde{r}_{\!\!_A}}$, as the IR cutoff ($L$). Using $L={\tilde{r}_{\!\!_A}}$ allows one to use the standard definition of horizon temperature, $T=1/(2\pi{\tilde{r}_{\!\!_A}})$, in eq.~(\ref{eq:FLT_DE}). Although there are several choices for entropies, here we will restrict ourselves to Bekenstein-Hawking, Barrow \cite{BARROW2020135643}, Tsallis \cite{Tsallis1988} and R\'{e}nyi \cite{renyi1961measures} entropies. Other entropies such as  Sharma -- Mittal entropy \cite{sharma1975new}, Kaniadakis entropy \cite{PhysRevE.72.036108} or Nojiri -- Odintsov -- Faraoni entropy \cite{PhysRevD.105.044042} represents different generalization schemes, which we will address in future works.

\section{Standard HDE versus Moradpour et al.'s proposal}
In this section we compare the Standard HDE with Moradpour et al.'s proposal. To compare, we consider two definitions for dark energy density, $\rho_{_{\Lambda 1}}$ from the CKN bound and $\rho_{_{\Lambda 2}}$ from Moradpour et al.'s proposal. Each of which respectively reads,
\begin{subequations}
	\begin{align}
	\rho_{_{\Lambda 1}}&=\mathcal{C}\frac{S}{{\tilde{r}_{\!\!_A}}^4}.\label{eq:Def1}\\
	\rho_{_{\Lambda 2}}&=\mathcal{C}_{_{\tiny T}}T\frac{dS}{dV}.\label{eq:Def2}
	\end{align}
\end{subequations}
In the previous section we discussed the motivation for the above expressions. The entropy choices to contrast the above definitions are (in natural units),
\begin{align}
S=\begin{cases}
S_{_\text{BH}} = \pi {\tilde{r}_{\!\!_A}}^2 & \text{: Bekenstein-Hawking,}\label{eq:Entropies}\\
S_{_\text{B }}~=\left(\pi {\tilde{r}_{\!\!_A}}^2\right)^{1+\frac{\Delta}{2}}&\text{: Barrow,}\\
S_{_\text{TC}}=\gamma(\pi{\tilde{r}_{\!\!_A}}^2)^{\delta}&\text{: Tsallis-Cirto,}\\
S_{_\text{R }}~=\frac{1}{\lambda}\log\left(1+\lambda\pi{\tilde{r}_{\!\!_A}}^2\right)&\text{: R\'{e}nyi.}
\end{cases}
\end{align}
Here, $\Delta$, $\delta$ and $\lambda$ are the non-extensive parameters for Barrow, Tsallis and R\'{e}nyi entropies respectively. Although some of the above entropies might look similar, the motivations behind each are different. For more details on different non-extensive entropies, one can consult references \cite{PhysRevD.105.044042, NOJIRI2022137189}  and the references therein. Here we have,

\paragraph{\textbf{Barrow entropy:}} In 2020, John Barrow considered fractal structure to the black hole horizon and introduced a toy model to account for the effects of quantum gravity on spacetime~\cite{BARROW2020135643}. Due to the fractal structure, the horizon that bounds a finite volume can have a finite or an infinite surface area. Here, $0\le\Delta\le1$ is the permissible range of the non-extensive parameter $\Delta$. When $\Delta=0$, we recover a smooth horizon that satisfies the standard area law. Although $\Delta$ can be a varying function \cite{Nojiri2019}, here we consider it as an unknown constant.

\paragraph{\textbf{Tsallis-Cirto entropy:}} One of the biggest difference between the conventional Boltzmann-Gibbs entropy and the Bekenstein-Hawking entropy is how they scale. While the former scales as $\sim L^3$, the latter scales like $\sim L^2$. In order to describe them in a unified framework, Tsallis and Cirto \cite{Tsallis2013} proposed a generalised version of Tsallis statistics \cite{Tsallis1988} to replace the Boltzmann-Gibbs statistics. The new generalised entropy reads,
\begin{equation}
S_{\beta,\delta}=\sum_{i=1}^{W}p_i\left[\log_{\beta}\left(\frac{1}{p_i}\right)\right]^{\delta},
\label{eq:basic_Tsallis}
\end{equation}
where $\beta$ is the Tsallis parameter, $\delta$ is the Tsallis-Cirto parameter, $W$ is the total number of internal configurations and $\{p_i\}$ is the probability distribution. Here $\log_{\beta}(x)\equiv \left(x^{1-\beta}-1\right)/(1-\beta)$. For $\beta=\delta=1$, the above expression is additive, and non-additive otherwise. 
The new composition rule reproduces the ordinary and the black hole entropy for different values of non-extensive parameters $\beta$ and $\delta$. In the context of black holes and cosmology, it takes the form given in equation (\ref{eq:Entropies}) as mentioned in \cite{Lymperis2018}. For a system that obeys the area law, Total number of internal configuration $W$ is a function of $L$ given as, $\ln W(L)\propto L^{d-1}, (d>1)$. Here $L$ is the length of the system such that $L^d$ is the volume in $d$-dimension. Then Tsallis and Cirto showed that, $S_{1,\delta=d/(d-1)}\propto S_{_\text{BH}}^{d/(d-1)}$, where $S_{_\text{BH}}$ is the Bekenstein-Hawking area law. In other words, $S_{_\text{TC}}=S_{1,\delta}\propto S_{_\text{BH}}^\delta$. Thus, in the context of cosmology we consider only a single parameter Tsallis-Cirto entropy by fixing the Tsallis parameter to unity. Here we assume $\delta$ as an unknown constant and $\gamma$ is a numerical factor introduced to make the expression dimensionally consistent in appropriate units (natural units in this case). Additionally, the value of $\gamma$ also depends on $\delta$. One may also assume a dynamic $\delta$, which we do not consider in this manuscript.

\paragraph{\textbf{R\'{e}nyi entropy:}} Another generalized version of entropy is the R\'{e}nyi entropy. It also appears as the formal logarithm of the Tsallis entropy, which allows one to confirm the compatibility of non-extensive thermodynamics and the zeroth law \cite{PhysRevE.83.061147}. Here, $\lambda$ is the non-extensive parameter introduced by R\'{e}nyi in \cite{renyi1961measures}. Unlike Tsallis-Cirto entropy discussed above, R\'enyi entropy is a single parameter generalization of the standard Boltzmann-Gibbs entropy. In the context of cosmology and black hole horizons, one can write the R\'enyi entropy as given in equation (\ref{eq:Entropies}). Czinner and Iguchi \cite{CZINNER2016306} introduced this expression to study the thermodynamic stability of black holes. This follows form the fact that the R\'enyi entropy is the formal logarithm of Tsallis entropy and Bekenstein-Hawking entropy is a special case of Tsallis-Cirto entropy. In the context of quantum information theory, R\'{e}nyi entropy is a suitable measure of entanglement between quantum states other than the standard von Neumann entropy. In the classical information theory, it is the generalization of the renowned Shannon entropy \cite{RevModPhys.81.865}. The definition of Tsallis-Cirto and R\'enyi entropy relies on the assumption that, $\ln W\propto$ Area is correct. And it is this proportionality with the area, instead of volume, that makes these entropies non-extensive.

Now, to compare equation (\ref{eq:Def1}) and (\ref{eq:Def2}), let us consider the simplest $3+1$ dimensional flat FLRW universe, where ${\tilde{r}_{\!\!_A}}=1/H$. Now using different choices of entropies given in (\ref{eq:Entropies}), we compute the corresponding expressions for dark energy densities, and we summarize the results in table (\ref{tab:1}). In the calculations, we assumed $T=1/(2\pi{\tilde{r}_{\!\!_A}})$ as the horizon temperature and $V=(4/3)\pi{\tilde{r}_{\!\!_A}}^3$ as the areal volume. 

\begin{table}[h]
	\centering
	\begin{tabular}{ccc}
		\hline
		Entropy& $\rho_{_{\Lambda 1}}$ & $\rho_{_{\Lambda 2}}$ \\
		\hline
		&  &  \\
		$\displaystyle\pi {\tilde{r}_{\!\!_A}}^2$& $\displaystyle\mathcal{C}\frac{\pi}{{\tilde{r}_{\!\!_A}}^2}$  &$ \displaystyle\frac{\mathcal{C}_{_{\tiny T}}}{4 \pi {\tilde{r}_{\!\!_A}}^{2}}$  \\
		&  & \\
		$\displaystyle\left(\pi {\tilde{r}_{\!\!_A}}^2\right)^{1+\frac{\Delta}{2}}$& $\displaystyle\mathcal{C}\frac{\left(\pi {\tilde{r}_{\!\!_A}}^2\right)^{1+\frac{\Delta}{2}}}{{\tilde{r}_{\!\!_A}}^4}$ & $\displaystyle \mathcal{C}_{_{\tiny T}}\frac{ \left(\pi {\tilde{r}_{\!\!_A}}^{2}\right)^{\frac{\Delta}{2} + 1} \left(\frac{\Delta}{2} + 1\right)}{4 \pi^{2} {\tilde{r}_{\!\!_A}}^{4}}$  \\
		&  &  \\
		$\displaystyle\gamma(\pi{\tilde{r}_{\!\!_A}}^2)^{\delta}$& $\displaystyle\mathcal{C}\frac{\gamma  \left(\pi {\tilde{r}_{\!\!_A}}^{2}\right)^{\delta}}{{\tilde{r}_{\!\!_A}}^{4}}$ & $\displaystyle\mathcal{C}_{_{\tiny T}}\frac{\delta \gamma  \left(\pi {\tilde{r}_{\!\!_A}}^{2}\right)^{\delta}}{4 \pi^{2} {\tilde{r}_{\!\!_A}}^{4}}$ \\
		&  &  \\
		$\displaystyle\frac{1}{\lambda}\log\left(1+\lambda\pi{\tilde{r}_{\!\!_A}}^2\right)$& $\displaystyle\frac{\mathcal{C}}{\lambda{\tilde{r}_{\!\!_A}}^4}\log\left(1+\lambda\pi{\tilde{r}_{\!\!_A}}^2\right)$ & $\displaystyle \mathcal{C}_{_{\tiny T}}\frac{1}{4 \pi {\tilde{r}_{\!\!_A}}^{2} \left(\pi \lambda {\tilde{r}_{\!\!_A}}^{2} + 1\right)}$  \\
		&  &  \\
		\hline
	\end{tabular}
	\caption{\label{tab:1}Different entropies and corresponding dark energy densities using the standard HDE density in equation (\ref{eq:Def1}) and Moradpour et al.'s proposal in equation (\ref{eq:Def2}).}
\end{table}

For a constant non-extensive parameter, one can easily see that both definitions appear to be proportional to each other for Bekenstein-Hawking, Barrow and Tsallis entropies. However, the case differs for R\'enyi entropy when $\lambda\neq0$. This result indicates that, although both definitions give similar scaling, they are not consistent with each other for all entropies. 

This inconsistency could arise due to many reasons, one of which might be the simple proportionality argument used in the relation (\ref{eq:Def2}). In the standard thermodynamic approach, the choice of entropy modifies the Friedmann equations, and this modification affects the relation between the deceleration parameter ($q$) and the total equation of state parameter ($w$). In the following sections, we will derive an expression for dark energy density in the late epoch from the laws of thermodynamics and show that, even in the simplest case of a flat $3+1$ dimensional FLRW universe, it is impossible to ensure the proportionality for all entropy choices.

Another reason for the above inconsistency might be the standard horizon temperature and volume in eq.~(\ref{eq:Def2}). In black hole thermodynamics, the Hawking temperature is, $T_{\text{\tiny H}}=\left({\partial S}/{\partial M}\right)^{-1}_{J,Q},$ where $M$, $J$ and $Q$ are the black hole's mass, angular momentum and electric charge respectively \cite{PhysRevD.15.2738}. If we modify the functional form of $S$ without modifying $M$, this relation will yield different expressions for $T_{\text{\tiny H}}$. Nojiri~et al.~\cite{PhysRevD.104.084030} noted this critical aspect and illustrated how the use of different entropies leads to inconsistencies in black hole thermodynamics. Thus, when we replace Bekenstein-Hawking entropy with other non-extensive entropies, even if $T$ has the dimension of $[{\tilde{r}_{\!\!_A}}^{-1}]$, it is not obvious that $T\propto1/{\tilde{r}_{\!\!_A}}$ is always valid. Further, if we follow the arguments by Cai and Kim \cite{Cai_2005} and assume that the temperature corresponding to the apparent horizon is $1/(2\pi {\tilde{r}_{\!\!_A}})$, we cannot claim that equations (\ref{eq:Def1}) and (\ref{eq:Def2}) represent the same quantity.

Furthermore, in the standard thermodynamic approach, one can only obtain the standard Friedmann equations starting from the Bekenstein-Hawking entropy. In the process, we will get an integration constant, which many authors consider as the cosmological constant \cite{Cai_2005}. Thus, in the thermodynamic approach, adding additional cosmic components to explain the accelerated expansion is generally not a trend. Restricted covariant theories also give the same explanation, where the cosmological constant depends on the initial conditions \cite{BUCHMULLER1988292}. 

Finally, the inconsistency might be due to all of the above reasons. Then, it is not trivial to bring a simple connection between holography and thermodynamics. The argument that any modified version of HDE density, either from equation (\ref{eq:Def1}) or from relations similar to equation (\ref{eq:Def2}), reducing to the standard HDE model with Hubble scale is not pressing enough, as such an HDE model cannot explain the late time acceleration. Instead, if an expression reduces to the $\Lambda$CDM model, that would be more convincing. 

To derive the expression for dark energy density from the laws of thermodynamics we need certain tools, which we will introduce here. For this purpose, we will assume a general $n+1$ dimensional FLRW spacetime with the metric given as,
\begin{equation}
ds^2=h_{ij}dx^idx^j+a^2(t)r^2d\Omega_{n-1}^2.
\end{equation}
Here, $i,j\in(0,1)$, $x^0=t$, $x^1=r$, $h_{00}=-1$, $h_{11}=a^2(t)/(1-kr^2)$, with scale factor `$a(t)$', comoving distance `$r$', $n-1$ spatial surface metric `$d\Omega_{n-1}$' and $h_{10}=h_{01}=0$. Further, $k$ can take values 1, 0 and -1 for closed, flat and open universes, respectively. It is convenient to discuss further in terms of distance $\tilde{r}=a(t)r$, as we are interested in a spherically symmetric apparent horizon which satisfies the condition $h_{ab}\partial_a\tilde{r}\partial_b\tilde{r}=0$. One can satisfy this condition for $\tilde{r}={\tilde{r}_{\!\!_A}}$ given as, 
\begin{equation}
{\tilde{r}_{\!\!_A}}=\left[{H^2+\frac{k}{a^2(t)}}\right]^{-1/2}. 
\label{eq:apparent_Horizon}
\end{equation}
Here, ${\tilde{r}_{\!\!_A}}$ is the radius of the apparent horizon. For a flat universe, with $k=0$, the above relation reduces to ${\tilde{r}_{\!\!_A}}=1/H$ with $H$ as the Hubble parameter. Additionally, $\Omega_n=2\pi^{n/2}/(n\Gamma(n/2))$,  such that, $A=n\Omega_n{\tilde{r}_{\!\!_A}}^{n-1}$ is the area of the apparent horizon. The temperature of apparent horizon, as motivated by black hole thermodynamics is, $T=\kappa/(2\pi)$, where, $\kappa~=~(2\sqrt{-h})^{-1}\partial_i(\sqrt{-h}h^{ij}\partial_j{\tilde{r}_{\!\!_A}})$ is the surface gravity. A straightforward calculation using the FLRW metric yields,
\begin{equation}
T=\frac{-1}{2\pi{\tilde{r}_{\!\!_A}}}\left(1-\frac{\dot{\tilde{r}}_{\!\!_A}}{2H{\tilde{r}_{\!\!_A}}}\right).
\label{eq:temperature}
\end{equation}
Using the above definitions, we can derive the energy density bound from the Clausius relation and the unified first law. Before that, let us see the problem with the proportionality argument used in \cite{Moradpour2018} when we start from the first law of thermodynamics.

\section{Is $\displaystyle \rho_{_{\Lambda }}\propto T\left(dS/dV\right)$ consistent with the first law?}

In this section, we will reconsider the simplest case of a flat $3+1$ dimensional FLRW universe and assume that the total energy density is $\rho\approx\rho_{_{\Lambda}}$. We then check whether the relation $\rho_{_{\Lambda }}\propto T\left(dS/dV\right)$ is consistent with the first law. Now from the Clausius relation we have, $-dE=TdS$ and the explicit expression reads \cite{Cai_2005},
\begin{equation}
4\pi{\tilde{r}_{\!\!_A}}^2(\rho+p)H{\tilde{r}_{\!\!_A}} dt=TdS.
\end{equation}
With $V=(4/3)\pi{\tilde{r}_{\!\!_A}}^3$, we have $dV=4\pi{\tilde{r}_{\!\!_A}}^2d{\tilde{r}_{\!\!_A}}$ and from ${\tilde{r}_{\!\!_A}}=1/H$, we have $d{\tilde{r}_{\!\!_A}}/dt = -\dot{H}/H^2\implies dt = (H^2/\dot{H})d{\tilde{r}_{\!\!_A}}$. Thus the above expression becomes,
\begin{equation}
\frac{-\left(\rho+p\right)H^2}{\dot{H}}dV=TdS.
\end{equation}
Since pressure $p:= w\rho$, we finally get,
\begin{equation}
\rho=\left[\frac{-\dot{H}}{(1+w)H^2}\right]T\frac{dS}{dV}.
\end{equation}
The above expression is an exact equation for the ``total'' energy density. Since we assumed a dark energy-dominated universe, we can follow Moradpour et al.'s argument and write,
\begin{equation}
\rho_{_{\Lambda }}\propto\left[\frac{-\dot{H}}{(1+w)H^2}\right]T\frac{dS}{dV}.
\end{equation}
Comparing the above expression with Moradpour et al.'s proposal in equation (\ref{eq:Def2}) demands,
\begin{equation}
\left[\frac{-\dot{H}}{(1+w)H^2}\right] = \text{a constant}.
\label{eq:condition1}
\end{equation}
Since, the deceleration parameter $q\equiv-1-\left(\dot{H}/H^2\right)$, the above constrain implies a specific class of relation between $q$ and $w$.

Now, to check the validity of Moradpour et al.'s proposal, let us first consider the standard Friedmann equations,
\begin{subequations}
	\begin{align}
	3H^2&={8\pi }\rho\label{eq:SFE1},\\
	3H^2+{2}\dot{H}&=-{8\pi }p\label{eq:SFE2}.
	\end{align}
\end{subequations}
Dividing the equation (\ref{eq:SFE2}) by equation (\ref{eq:SFE1}) we get,
\begin{align}
&1+\frac{2}{3}\left(\frac{\dot{H}}{H^2}\right)=-w\implies 1+w=-\frac{2}{3}\left(\frac{\dot{H}}{H^2}\right)\nonumber\\&\implies\frac{\dot{H}}{(1+w)H^2}=-\frac{3}{2}.
\label{eq:19}
\end{align}
Thus, for Bekenstein-Hawking entropy, Moradpour et al.'s proposal appear to be valid. We again emphasise that, in the thermodynamic approach, one can only derive the standard Friedmann equations using the Bekenstein-Hawking entropy. 

Let us now consider other non-extensive entropies. Since Barrow and Tsallis-Cirto entropies are mathematically similar, we will consider Barrow-modified Friedmann equations. With the same background assumptions, the Barrow corrected Friedmann equations read \cite{PhysRevD.103.123503} (in natural units),
\begin{subequations}
	\begin{align}
	&3H^{2-\Delta} = \frac{{8\pi} \rho}{\pi^{\Delta/2}} \left(
	\frac{2-\Delta}{2+\Delta}\right)\label{eq:BFE1},\\
	&(2-\Delta)\frac{\ddot{a}}{a}
	H^{-\Delta}+(1+\Delta)H^{2-\Delta}=-\frac{{8\pi} p}{\pi^{\Delta/2}} \left(
	\frac{2-\Delta}{2+\Delta}\right)\label{eq:BFE2}.
	\end{align}
\end{subequations}
With $\ddot{a}=\dot{H}+H^2$, one can easily show that, 
\begin{equation}
\frac{\dot{H}}{(1+w)H^2}=-\frac{3}{2-\Delta}.
\end{equation}
Since we started with a constant $\Delta$, this again is a constant. Hence, for Barrow and Tsallis-Cirto entropies,  Moradpour et al.'s proposal appear to be valid. This is why we get similar expressions for dark energy density from equation (\ref{eq:Def2}) and eq.~(\ref{eq:Def1}). (See table (\ref{tab:1})).

\begin{figure}[h]
	\centering
	\includegraphics[width=0.45\textwidth]{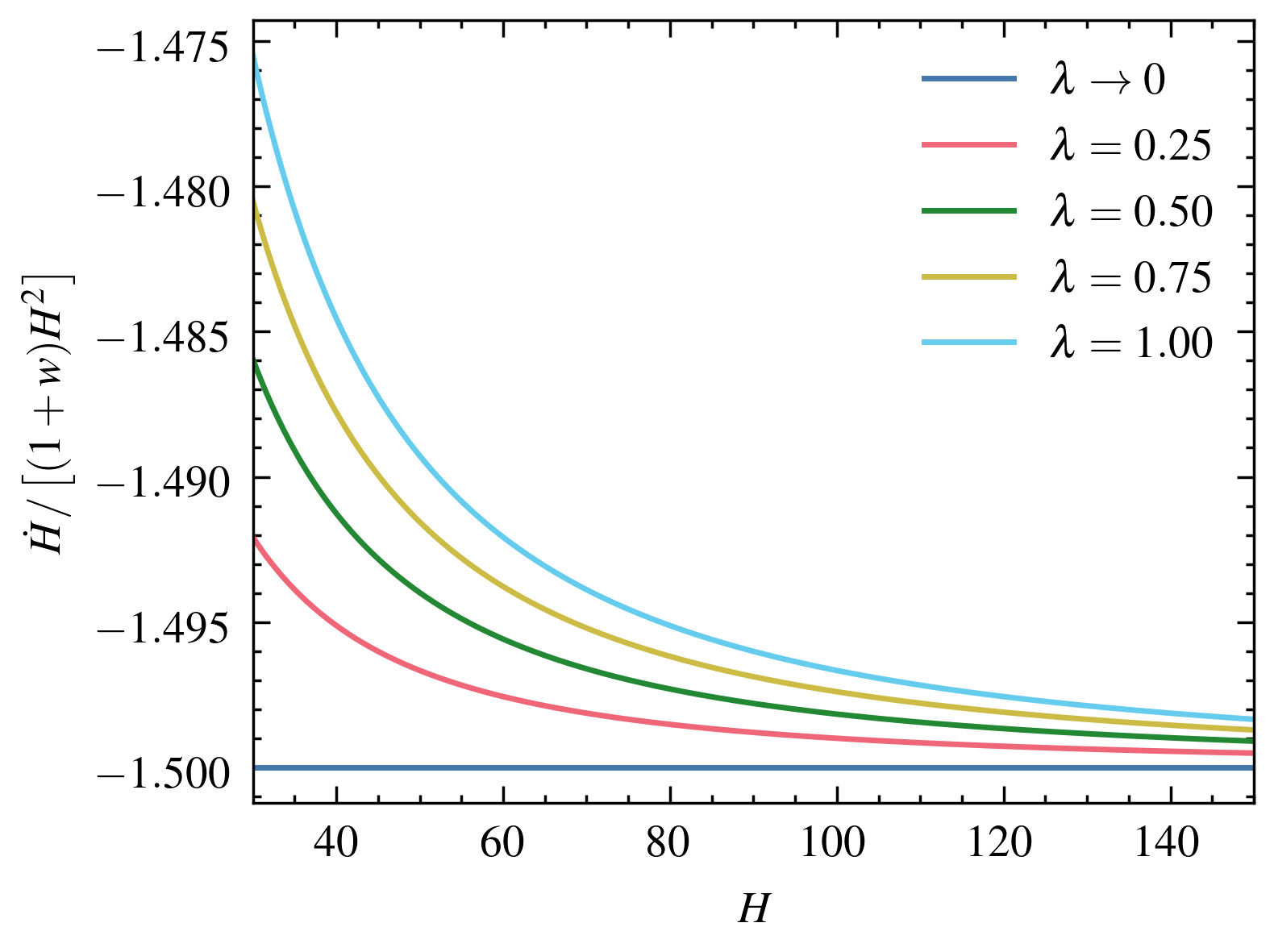}
	\caption{${\dot{H}}/\left[{(1+w)H^2}\right]$ vs $H$ in R\'enyi modified Friedmann equations for different values of $\lambda$.}
	\label{fig:1}
\end{figure}

In the same manner, considering the R\'{e}nyi entropy, we get the corresponding R\'enyi modified Friedmann equations as, (See \ref{AppendixA}.)
\begin{subequations}
	\begin{align}
	&3H^2+3\lambda\pi\log\left(1+\frac{H^2}{\pi\lambda}\right)=8\pi  \rho\label{eq:RFE1},\\
	&\frac{2 H^{2} \dot{H}}{\left(H^{2} + \pi \lambda\right)}+3H^2+3\lambda\pi\log\left(1+\frac{H^2}{\pi\lambda}\right)=-8\pi p.
	\label{eq:RFE2}
	\end{align}
\end{subequations} 
In this case, we get,
\begin{align}
&\frac{\dot{H}}{(1+w)H^2}=\nonumber\\&-\frac{3}{2} \frac{ \left(H^{2} + \pi \lambda\right) \left[H^{2} + \pi \lambda \log{\left(1+\frac{H^2}{\pi\lambda} \right)} \right]}{ H^{4}}.
\end{align}
which is a function of $H$ for non-zero $\lambda$. Figure~(\ref{fig:1}) illustrates the functional behaviour of ${\dot{H}}/({(1+w)H^2})$ for different values of $\lambda$.  In the limit $\lambda\rightarrow0$, equations (\ref{eq:RFE1}) and (\ref{eq:RFE2}) reduces to the standard Friedmann equations, and we recover the condition in equation (\ref{eq:condition1}). 
Thus, according to the first law of thermodynamics, the proposal in ref. \cite{Moradpour2018} is only consistent with the standard HDE when we consider the standard Friedmann equations or entropies that follow exponent stretched area laws. One can also prove this by equating equation (\ref{eq:Def1}) and (\ref{eq:Def2}), and then solve for $S$. In doing so, we get,
\begin{equation}
S=\mathcal{K}{\tilde{r}_{\!\!_A}}^{\frac{8\pi^2\mathcal{C}}{\mathcal{C}_{_{\tiny T}}}}.
\end{equation}
Here, $\mathcal{K}$ accounts for the integration constant. When ${8\pi^2\mathcal{C}}/{\mathcal{C}_{_{\tiny T}}}=2$, we recover the standard area law, and any deviation might point towards Barrow or Tsallis-Cirto type area stretched entropies. This is why we say that the simple proportionality argument used in \cite{Moradpour2018} may not be generally valid. The other possibility is to redefine the horizon temperature or the areal volume as in \cite{PhysRevD.104.084030}. Redefining the temperature essentially redefines the surface gravity and thus the underlying metric. Since that would lead to a universe different from what we expect from an FLRW spacetime, redefining the temperature appears premature. 

Till now, we considered the first law of thermodynamics and the Friedmann equations (standard/modified). However, one needs to consider the continuity relation to get the Friedmann equations from the first law. In the following subsections, we use the continuity equation and derive an expression for energy density without explicitly using the standard (modified) Friedmann equations and study whether it can resolve the inconsistency we encountered earlier. For simplicity, we do not consider any interaction between dark energy and matter. It will be interesting to investigate such situations, which we will consider elsewhere. 

\section{Dark energy density from the Clausius relation}

In the previous section we checked whether $\rho_{_{\Lambda }}\propto T\left(dS/dV\right)$ is consistent with the first law or not. It is possible to reverse the question and ask ``\textit{what form of $\rho_{_{\Lambda }}$ one can obtain from the first law?}''. In this section we derive an expression for the dark energy density from the Clausius relation. One of the seminal works that first illustrated the deep connection between gravity and thermodynamics belongs to Jacobson \cite{PhysRevLett.75.1260}, where he obtained Einstein's field equation from the Clausius relation, $\delta Q=TdS$. In cosmology, we can take  $\delta Q=-dE$ as the heat flow across the apparent horizon, and the Clausius relation reads \cite{Cai_2005},
\begin{equation}
-dE=TdS,
\label{eq:ClausiusRelation}
\end{equation}
where, $-dE=A(\rho+p)H{\tilde{r}_{\!\!_A}} dt$, with area $A=n\Omega_n{\tilde{r}_{\!\!_A}}^{n-1}$ and $\Omega_n={\pi^{n/2}}/{\Gamma\left(\frac{n}{2}+1\right)}$. Now we assume the validity of the general continuity equation given as,
\begin{equation}
\dot{\rho}=-nH(\rho+p).
\label{eq:continutity_N}
\end{equation}
Here, $\rho$ is the total energy density, $p$ is the pressure, $H$ is the Hubble parameter, $n$ is the number of spatial dimensions, and the over-dot represents the derivative with respect to cosmic time. One may also write this relation using the equation of state parameter defined as $w:= p/\rho$. Now, using the expressions for $dE$, $A$ and the continuity equation, we arrive at,
\begin{align}
A(\rho+p)H{\tilde{r}_{\!\!_A}} dt&=n\Omega_n{\tilde{r}_{\!\!_A}}^{n-1}\left(\frac{-\dot{p}}{nH}\right)H{\tilde{r}_{\!\!_A}} dt=-\Omega_n{\tilde{r}_{\!\!_A}}^nd\rho.
\end{align}
Plugging this expression back to the Clausius relation given in equation (\ref{eq:ClausiusRelation}), we get,
\begin{align}
-\Omega_n{\tilde{r}_{\!\!_A}}^nd\rho=TdS
\implies \rho=-\int\frac{1}{\Omega_n{\tilde{r}_{\!\!_A}}^n}TdS.
\end{align}
Now we approximate the expression for $T$ in equation (\ref{eq:temperature}) as,
\begin{equation}
T\simeq\frac{1}{2\pi{\tilde{r}_{\!\!_A}}}.
\end{equation}
Although this is an approximation, it is valid, as the first law holds the same for an infinitesimal change in temperature. Thus we get the total energy density as
\begin{equation}
\rho=\frac{-1}{2\pi\Omega_n}\int\frac{1}{{\tilde{r}_{\!\!_A}}^{n+1}}dS. 
\end{equation}
We will show that the above result remains unchanged even when considering the unified first law, where we do not take the above approximation in temperature.

Now, the total energy density ($\rho$) consist of different cosmic components such as matter (including dark matter) ($\rho_{m}$), radiation ($\rho_{r}$), dark energy ($\rho_{_{\Lambda}}$) etc. In this article, we restrict ourselves to a period around transition redshift to late acceleration. Since $\rho_{r}$ dilutes much faster than any other components, we may ignore its effect in our analysis and assume the relation $\rho\simeq\rho_{_{\Lambda}}+\rho_{m}$ for an almost flat universe. Then the dark energy density is,
\begin{equation}
\rho_{_{\Lambda}}=\left(\frac{-1}{2\pi\Omega_n}\int\frac{1}{{\tilde{r}_{\!\!_A}}^{n+1}}dS\right)-\rho_m.
\end{equation}
In the asymptotic de Sitter limit, when redshift $z\rightarrow-1$, the matter density, which scales as $\rho_{m}\propto(1+z)^3$, dilutes away, and we get a ``limit'' where,
\begin{equation}
\rho_{_{\Lambda}}\rightarrow\frac{-1}{2\pi\Omega_n}\int\frac{1}{{\tilde{r}_{\!\!_A}}^{n+1}}dS.
\label{eq:genCKN1}
\end{equation}
In \cite{Moradpour2018}, the authors considered this bound as the source for HDE. Then the problem is that this expression is valid only in the limit where no matter component exists. One may also assume such a limit for a matter-dominated era, and we will end up in a cyclic loop of arguments \cite{HSU200413}. However, if we demand to define a thermodynamic analogue of HDE density from this limit as in \cite{Moradpour2018}, it can be of the form,
\begin{equation}
\rho_{_{\Lambda}}=\frac{-{C}^2}{2\pi\Omega_n}\int\frac{1}{{\tilde{r}_{\!\!_A}}^{n+1}}dS.
\label{eq:ThermoHDECR}
\end{equation}
Here, $C^2$ is the proportionality constant. We will now derive the same from the unified first law for completeness and then discuss its features.

\section{Dark energy density from the unified first law}

Similar to the previous section, here we derive a possible expression for dark energy density from the unified first law. The notable work by Akbar and Cai \cite{PhysRevD.75.084003} used the unified first law to discuss Einstein, Gauss-Bonnet and Lovelock gravities in a unified framework. In the previous section, we used an approximation in the temperature for our calculations, while in the unified first law,  there is no need to take such assumptions. Here, the unified first law of thermodynamics reads, 
\begin{equation}
dE=TdS+WdV,
\label{eq:UFL}
\end{equation}
where, $V=n\Omega_n$ is the volume, $W=\frac{1}{2}(\rho-p)$ is the work density and $E=\rho V$ is the total gravitating energy bounded by the horizon \cite{PhysRev.136.B571}.\footnote{ Reference \cite{PhysRevD.80.104016} gives a detailed account of the Misner-Sharp energy ($\rho V$), and references therein point out various other energy candidates.} Substituting for $E$ in equation (\ref{eq:UFL}) we get,
\begin{align}
&Vd\rho+\rho dV=TdS+\frac{1}{2}(\rho-p)dV
\nonumber\\&	\implies Vd\rho+\frac{1}{2}(\rho+p)dV=TdS.
\label{eq:unified_first_law}
\end{align}
Now, using the continuity equation given in equation (\ref{eq:continutity_N}) and $dV=n\Omega_n{\tilde{r}_{\!\!_A}}^{n-1}d{\tilde{r}_{\!\!_A}}$ to equation (\ref{eq:unified_first_law}), we arrive at,
\begin{align}
\Omega_n{\tilde{r}_{\!\!_A}}^n\left(1-\frac{\dot{\tilde{r}}_{\!\!_A}}{2H{\tilde{r}_{\!\!_A}}}\right)d\rho=TdS.
\end{align}
Now the proper definition of temperature in terms of surface gravity, $T=\kappa/(2\pi)$ reads,
\begin{equation}
T=\frac{-1}{2\pi{\tilde{r}_{\!\!_A}}}\left(1-\frac{\dot{\tilde{r}}_{\!\!_A}}{2H{\tilde{r}_{\!\!_A}}}\right).
\end{equation}
Then immediately, we get,
\begin{align}
&	\Omega_n{\tilde{r}_{\!\!_A}}^n\left(1-\frac{\dot{\tilde{r}}_{\!\!_A}}{2H{\tilde{r}_{\!\!_A}}}\right)d\rho=\frac{-1}{2\pi{\tilde{r}_{\!\!_A}}}\left(1-\frac{\dot{\tilde{r}}_{\!\!_A}}{2H{\tilde{r}_{\!\!_A}}}\right)dS
\nonumber\\&\implies\rho=\frac{-1}{2\pi\Omega_n}\int\frac{1}{{\tilde{r}_{\!\!_A}}^{n+1}}dS.
\end{align}
Once again, in the asymptotic de Sitter limit, when redshift $z\rightarrow-1$, the matter density, which scales as $\rho_{m}\propto(1+z)^3$, dilutes away, and we get the limit,
\begin{equation}
\rho_{_{\Lambda}}\rightarrow\frac{-1}{2\pi\Omega_n}\int\frac{1}{{\tilde{r}_{\!\!_A}}^{n+1}}dS.
\label{eq:genCKN2}
\end{equation}
This is the same as the previous result from the Clausius relation. Either way, we do get the same bound for the dark energy density as proposed by Moradpour et al. If we consider equation (\ref{eq:ThermoHDECR}) as our definition for HDE density from thermodynamics, we get expressions similar to $\rho_{_{\Lambda 2}}$ given in table (\ref{tab:1}), except for R\'enyi entropy. Table (\ref{tab:2}) summarise the expressions for dark energy density using equation (\ref{eq:ThermoHDECR}). It is again evident that equation (\ref{eq:ThermoHDECR}) generates HDE proportional to the HDE from the CKN bound, for Bekenstein-Hawking, Barrow and Tsallis entropies. However, similar to Moradpour et al.'s results, R\'enyi entropy gives a different result. 
\begin{table}[h]
	\centering
	\begin{tabular}{cc}
		\hline
		Entropy & $\displaystyle\rho_{_{\Lambda}}$ using equation (\ref{eq:ThermoHDECR}) \\
		\hline
		&\\
		$\displaystyle\pi {\tilde{r}_{\!\!_A}}^2$& $\displaystyle\frac{3C^2}{8\pi{\tilde{r}_{\!\!_A}}^2}$ \\
		&  \\
		$\displaystyle\left(\pi {\tilde{r}_{\!\!_A}}^2\right)^{1+\frac{\Delta}{2}}$ & $\displaystyle \frac{3 C^{2}}{8 \pi^{2}} \left(\begin{cases} \frac{\pi^{\frac{\Delta}{2} + 1} \left(\Delta + 2\right) \left({\tilde{r}_{\!\!_A}}^{2}\right)^{\frac{\Delta}{2}}}{{\tilde{r}_{\!\!_A}}^{2} \left(2-\Delta\right)} & \text{;}\: \Delta \neq 2 \\\pi^{2} \left(\Delta + 2\right) \log{\left(\frac{1}{{\tilde{r}_{\!\!_A}}} \right)} & \text{;}\: \Delta = 2 \end{cases}\right)$  \\
		&  \\
		$\displaystyle\gamma(\pi{\tilde{r}_{\!\!_A}}^2)^{\delta}$& $\displaystyle  \frac{3 C^{2} \delta \gamma}{4 \pi^{2}} \left(\begin{cases} \frac{\pi^{\delta} \left({\tilde{r}_{\!\!_A}}^{2}\right)^{\delta}}{4 {\tilde{r}_{\!\!_A}}^{4}-2 \delta {\tilde{r}_{\!\!_A}}^{4}} & \text{;}\: \delta \neq 2 \\\pi^{2} \log{\left(\frac{1}{{\tilde{r}_{\!\!_A}}} \right)} & \text{;}\: \delta = 2 \end{cases}\right)$\\
		&  \\
		$\displaystyle\frac{1}{\lambda}\log\left(1+\lambda\pi{\tilde{r}_{\!\!_A}}^2\right)$ & $ \displaystyle\frac{3 C^{2} }{8 \pi^{2}}\left[\frac{\pi}{{\tilde{r}_{\!\!_A}}^2} -\pi^2\lambda\log\left(1+\frac{1}{\pi\lambda{\tilde{r}_{\!\!_A}}^2}\right)\right]$ \\
		&   \\
		\hline
	\end{tabular}
	\caption{Different entropies and corresponding dark energy densities using the equation (\ref{eq:ThermoHDECR}).}
	\label{tab:2}
\end{table}

Interestingly, for R\'enyi entropy, equation (\ref{eq:ThermoHDECR}) is also different from Moradpour et al.'s definition given in equation (\ref{eq:Def2}). Why should the laws of thermodynamics give distinct expressions for the same quantity? This could be because, we considered only the validity of generalized continuity equation, while equation (\ref{eq:Def2}) demanded the validity of equation (\ref{eq:condition1}). However, one cannot guarantee the validity of equation (\ref{eq:condition1}) from the Clausius relation or the unified first law of thermodynamics. Further, it is not immediately clear from the arguments in \cite{Moradpour2018}, how one can assume $E_{_\Lambda}\sim\rho_{_{\Lambda }}V$ (the Misner-Sharp energy) and consider $dE_{_\Lambda}=\rho_{_{\Lambda }}dV$.

As already mentioned, standard Friedmann equations need Bekenstein-Hawking entropy as the entropy choice. Then, it is unclear how one can choose one entropy to define the equations of motion and another to define dark energy density from the same governing laws. For example, it is not logical to construct modified Friedmann equations using R\'enyi entropy and use Barrow entropy to define the dark energy density from the same laws of thermodynamics. A possible justification might be that the equations of motion come from Einstein's field equation and the expression for dark energy density has its roots in holography or thermodynamics. However, in the thermodynamic approach, Einstein's gravity, which yields the standard Friedmann equations, is only a particular case when we use Bekenstein-Hawking entropy. This leads us to investigate the possibility of proposing a dark energy component using the laws of thermodynamics and the notion of holography. 

\section{Standard R\'enyi HDE and its inconsistencies}

Before we jump into a new definition, let us see some inconsistencies in the standard approach to HDE, especially with R\'enyi entropy. Here, we consider the R\'enyi entropy and propose R\'enyi HDE density using the equation (\ref{eq:Def1}). It is interesting to investigate this, as there is no such work in the literature to our knowledge that uses equation (\ref{eq:Def1}) for R\'enyi HDE density. This might be because of the lack of simple analytical solutions for the Hubble parameter. Here the R\'enyi HDE density from the CKN bound reads,
\begin{equation}
\rho_{_{\Lambda}}=\frac{\mathcal{C}}{\lambda{\tilde{r}_{\!\!_A}}^4}\log\left(1+\lambda\pi{\tilde{r}_{\!\!_A}}^2\right).
\label{eq:RHDE_sDE}
\end{equation}
We will consider the Hubble horizon as the IR cutoff for our discussion. Now, the standard Friedmann equations are,
\begin{subequations}
	\begin{align}
	3H^2&={8\pi }(\rho_m+\rho_{_{\Lambda }})\text{ and }	3H^2+{2}\dot{H}=-{8\pi }p.
	\end{align}
\end{subequations}
For simplicity, in our analysis, we do not consider any interaction between HDE and matter. Thus each component obeys independent continuity equations, and for the matter sector, we get,
\begin{equation}
\dot{\rho}_m+3H\rho_m=0\implies\rho_m=\rho_{m0}(1+z)^3.
\end{equation}
Here $z$ is the redshift. Since it is convenient to deal with density parameters rather than actual densities we use,
\begin{subequations}
	\begin{align}
	&\Omega_m = \frac{8\pi\rho_m}{3H^2}\text{ and } \Omega_{_\Lambda} = \frac{8\pi\rho_{_\Lambda}}{3H^2} \text{, such that, }
	\Omega_m +\Omega_{_\Lambda}=1.
	\end{align}
\end{subequations}
Thus, using the expression for dark energy density, matter density and $H(z)=H_0h(z)$, we get,
\begin{align}
h^2(z)&=\Omega_{m0}(1+z)^3+\frac{8\pi\mathcal{C}(H_0h(z))^4}{3\lambda H_0^2}\log\left(1+\frac{\lambda\pi}{(H_0h(z))^2}\right).
\end{align}
where $\Omega_{m0}$ and $H_0$ are the present value (where $z=0$) of the matter density parameter and Hubble parameter. Now to obtain the expression for the free parameter $\mathcal{C}$, we consider the situation when $z=0$ for which we have $h(z=0)=1$. Then we get,
\begin{align}
1&=\Omega_{m0}+\frac{8\pi\mathcal{C}H_0^2}{3\lambda }\log\left(1+\frac{\lambda\pi}{H_0^2}\right)
\nonumber\\&	\implies \mathcal{C}=\frac{3\lambda\left(1-\Omega_{m0}\right)}{8\pi H_0^2\log\left(1+\frac{\lambda\pi}{H_0^2}\right)}.
\end{align}
Now the previous expression in terms of $H=H(z)$ becomes,
\begin{align}
H^2&=H_0^2\Omega_{m0}(1+z)^3+\frac{{\left(1-\Omega_{m0}\right)}H^4}{ H_0^2\log\left(1+\frac{\lambda\pi}{H_0^2}\right)}\log\left(1+\frac{\lambda\pi}{H^2}\right).
\end{align}
This equation makes it difficult to get an exact analytical form for $H$ as a function of~$z$. Thus, we resort to numerical methods to further investigate the model. It is noteworthy to recognize Moradpour et al.'s results, as it is possible to get an exact expression for $H$ in terms of $z$ from their proposal.

\begin{figure}[h]
	\centering
	\includegraphics[width=0.45\textwidth]{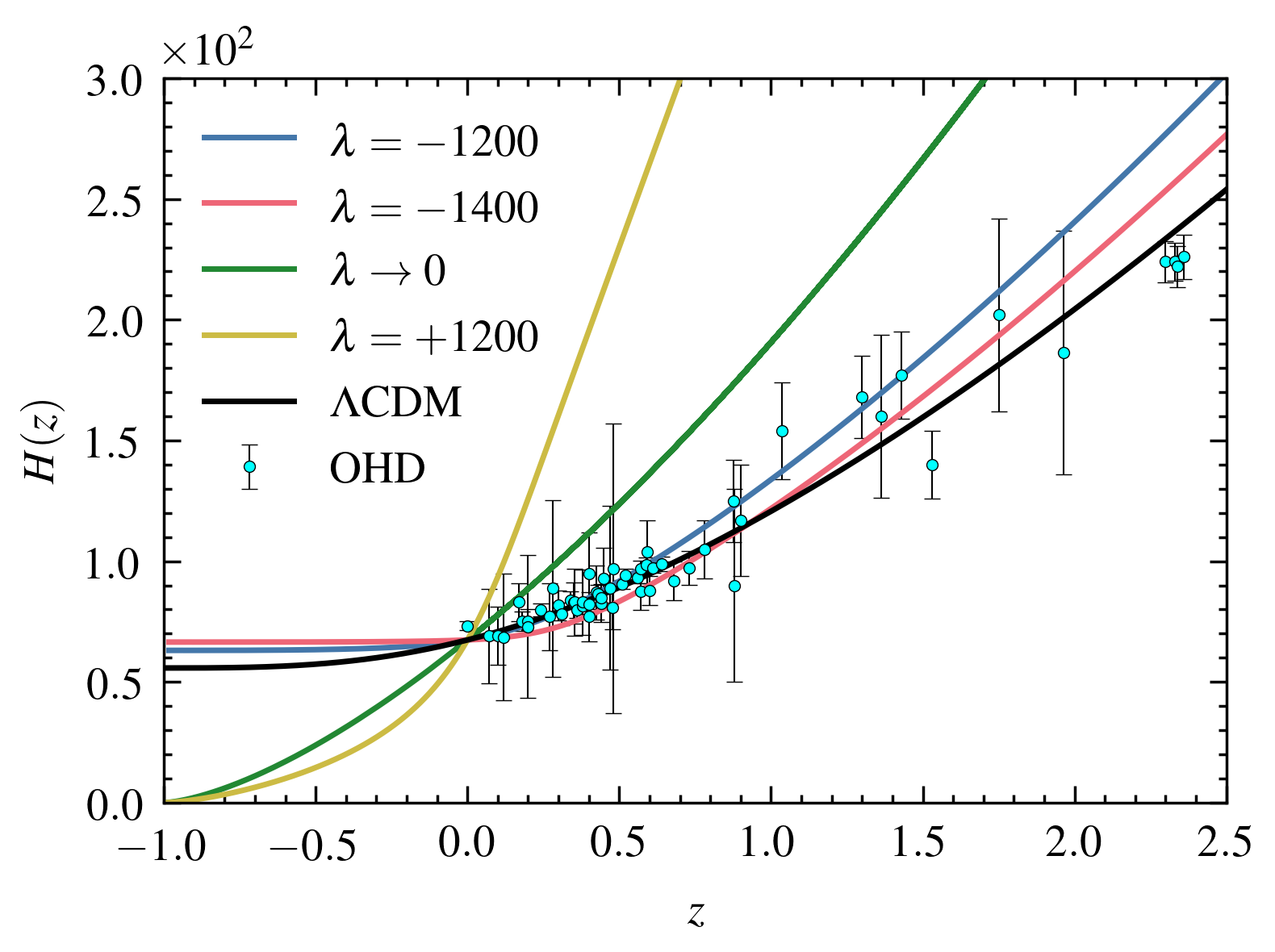}
	\caption{Hubble parameter ($H(z)$) as a function of redshift ($z$) for different values of $\lambda$ with standard R\'enyi HDE density (with $\Omega_{m0}=0.315$ and $H_0=67.4$ km/s/Mpc \cite{aghanim2020planck}). OHD corresponds to the Observational Hubble Data set with 57 entries \cite{https://doi.org/10.1002/asna.20220003}}
	\label{fig:ModelAData}
\end{figure}

In figure. (\ref{fig:ModelAData}), we plot the evolution of the Hubble parameter as a function of redshift for different values of $\lambda$. The model does not exhibit a late time acceleration for zero or large positive values of $\lambda$. However, similar to Moradpour et al.'s results, we get a late time acceleration for negative values of $\lambda$, which could explain the observation. The limit in which $\lambda\rightarrow0$, the expression reduces to ordinary HDE model and cannot explain the cosmic evolution, as we did not consider any interactions. 

\begin{figure}[h]
	\centering
	\includegraphics[width=0.45\textwidth]{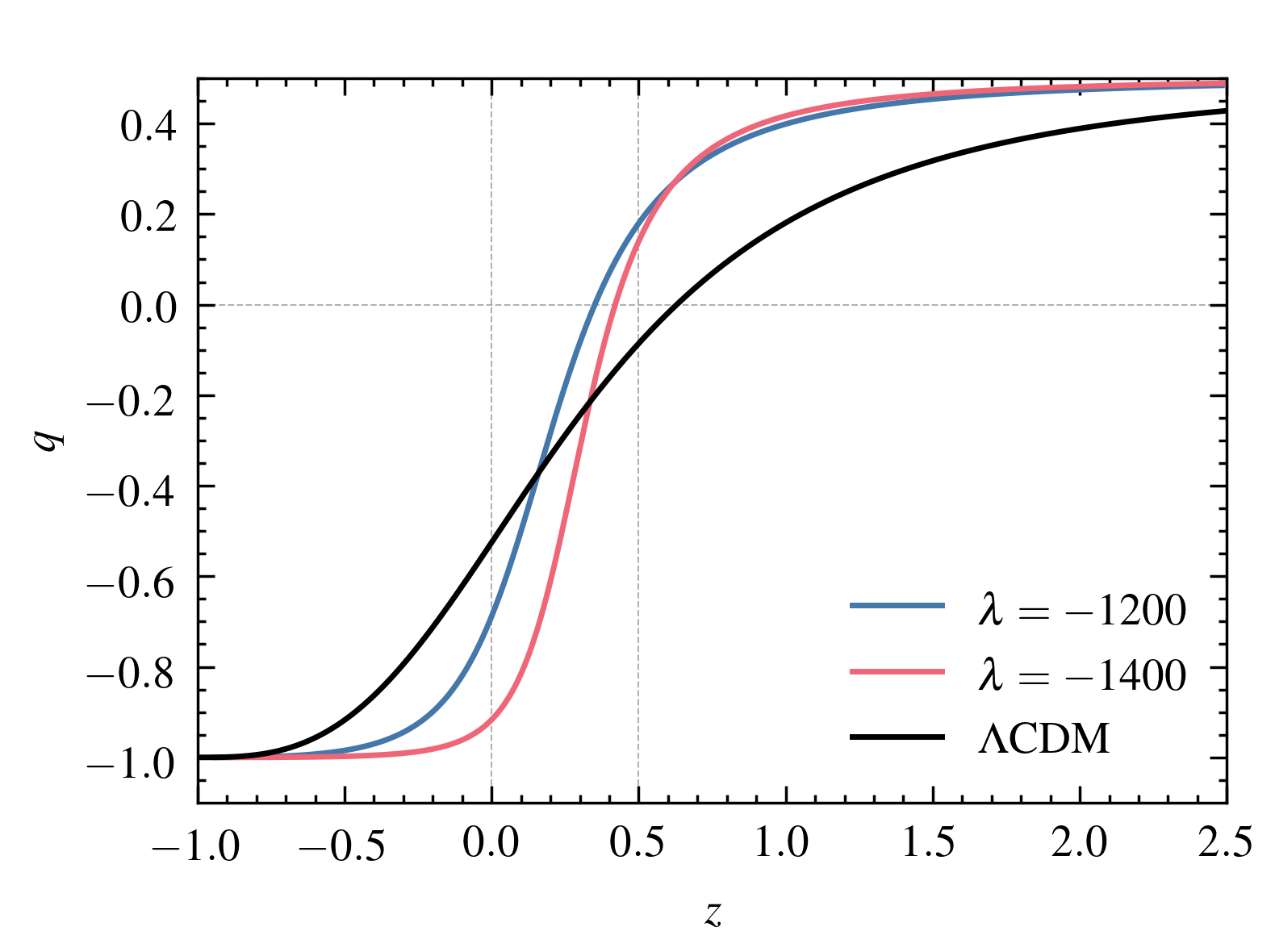}
	\caption{Deceleration parameter ($q$) as a function of redshift ($z$) for different values of $\lambda$ for standard R\'enyi HDE model (with $\Omega_{m0}=0.315$ and $H_0=67.4$ km/s/Mpc \cite{aghanim2020planck}).}
	\label{fig:ModelAq}
\end{figure}

To study the features of the cosmic evolution, we now investigate the deceleration parameter $q$ for $\lambda$ that closely follows the data points. The definition of $q$ is,
\begin{equation}
q\equiv-1-\frac{\dot{H}}{H^2}=-1+\frac{1+z}{H(z)}\left(\frac{dH(z)}{dz}\right)
\end{equation}
Evaluating the previous expression, we plot the evolution for $q$ as a function of redshift in figure (\ref{fig:ModelAq}) for different values of $\lambda$. The model shows late time acceleration with transition redshift close to $z=0.5$. Similar to $\Lambda$CDM, this model also exhibits a final de Sitter limit since $q\rightarrow-1$ implies $\dot{H}\rightarrow0$, which then implies a constant Hubble parameter.

\begin{figure}[h]
	\centering
	\includegraphics[width=0.45\textwidth]{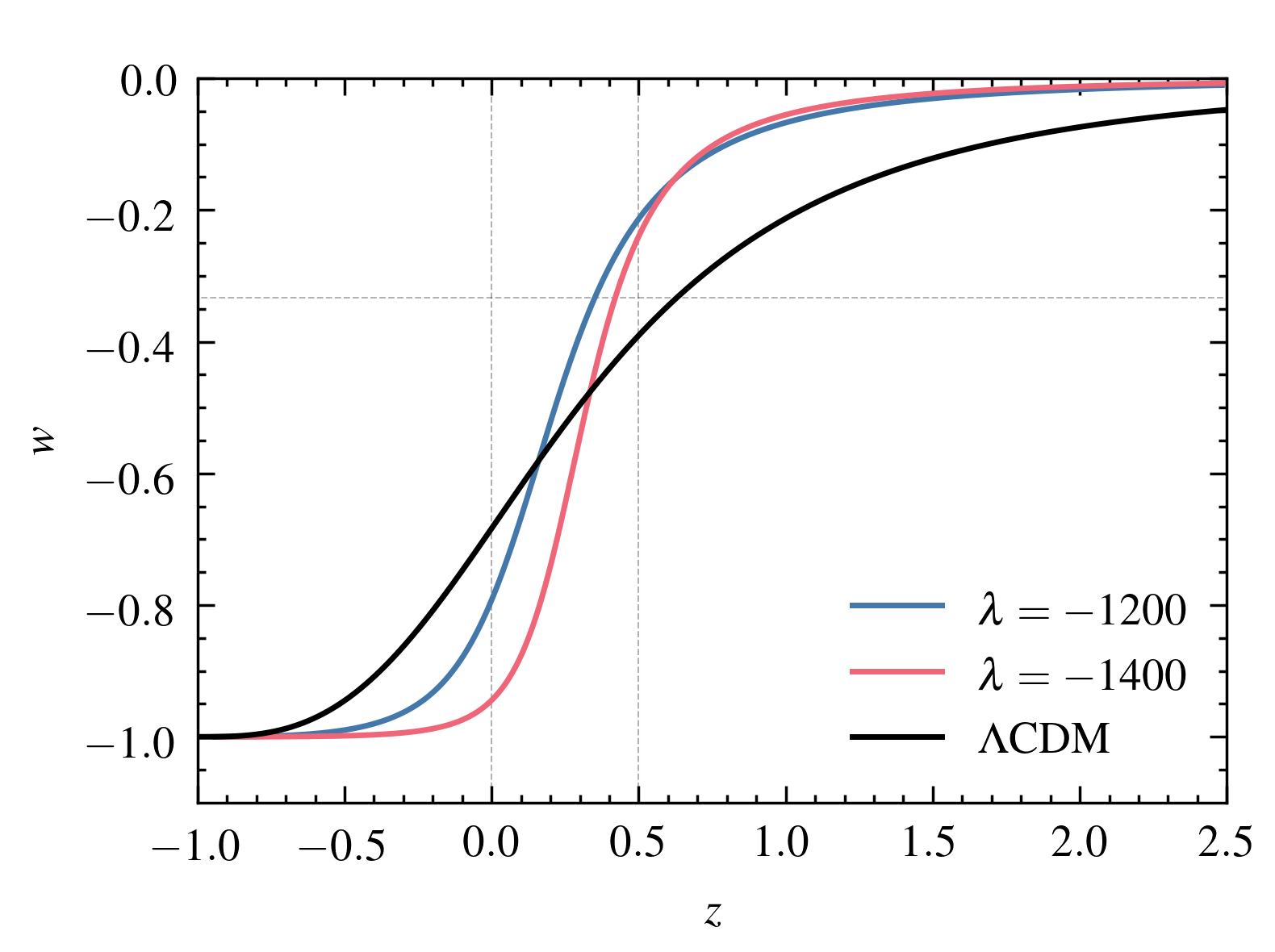}
	\includegraphics[width=0.45\textwidth]{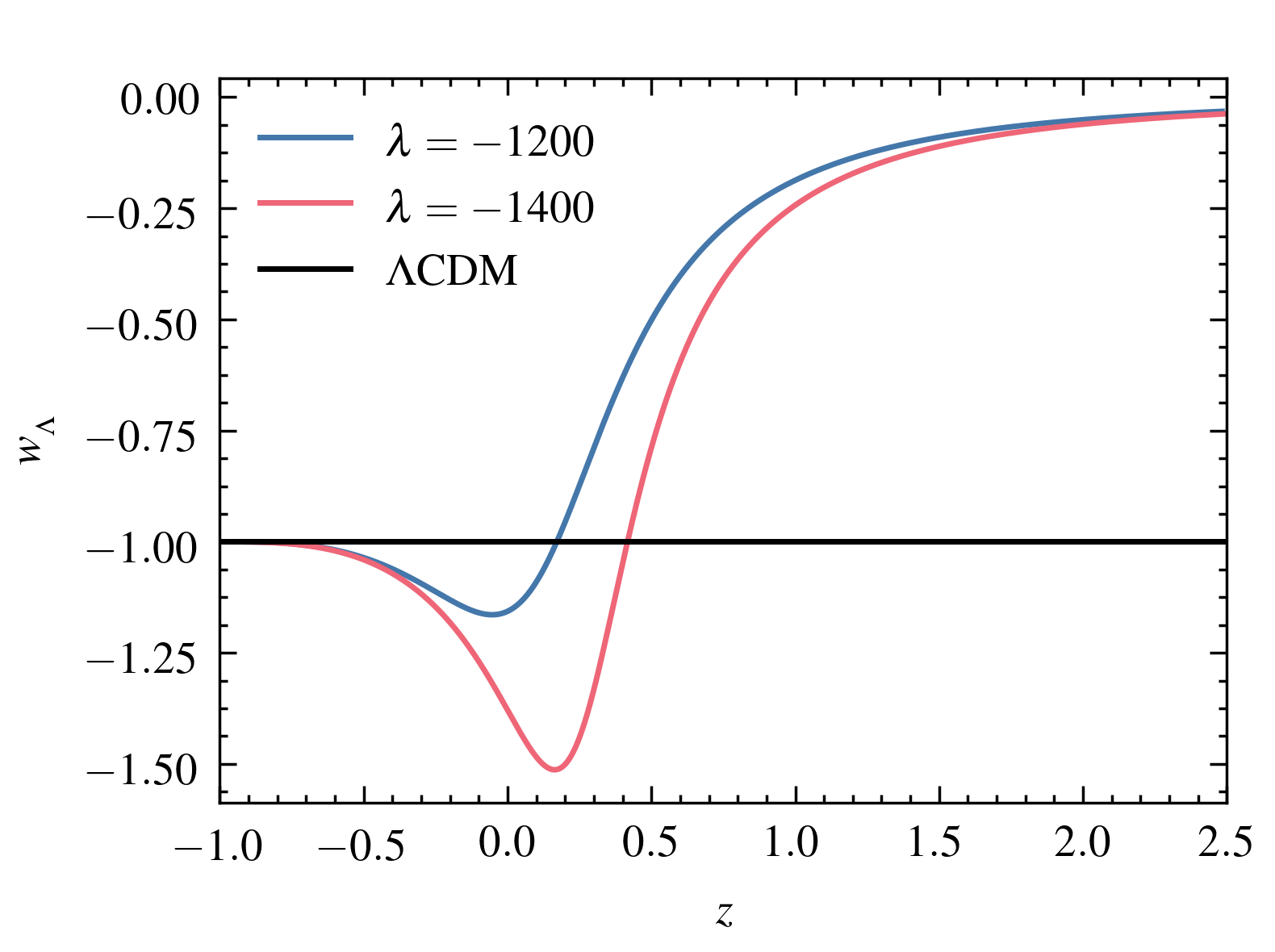}
	\caption{Total equation of state ($w$) and dark energy equation of state ($w_{_\Lambda}$) as a function of redshift ($z$) for different values of $\lambda$ with standard R\'enyi HDE density (with $\Omega_{m0}=0.315$ and $H_0=67.4$ km/s/Mpc \cite{aghanim2020planck}).}
	\label{fig:ModelAw}
\end{figure}

\begin{figure}[h]
	\centering
	\includegraphics[width=0.45\textwidth]{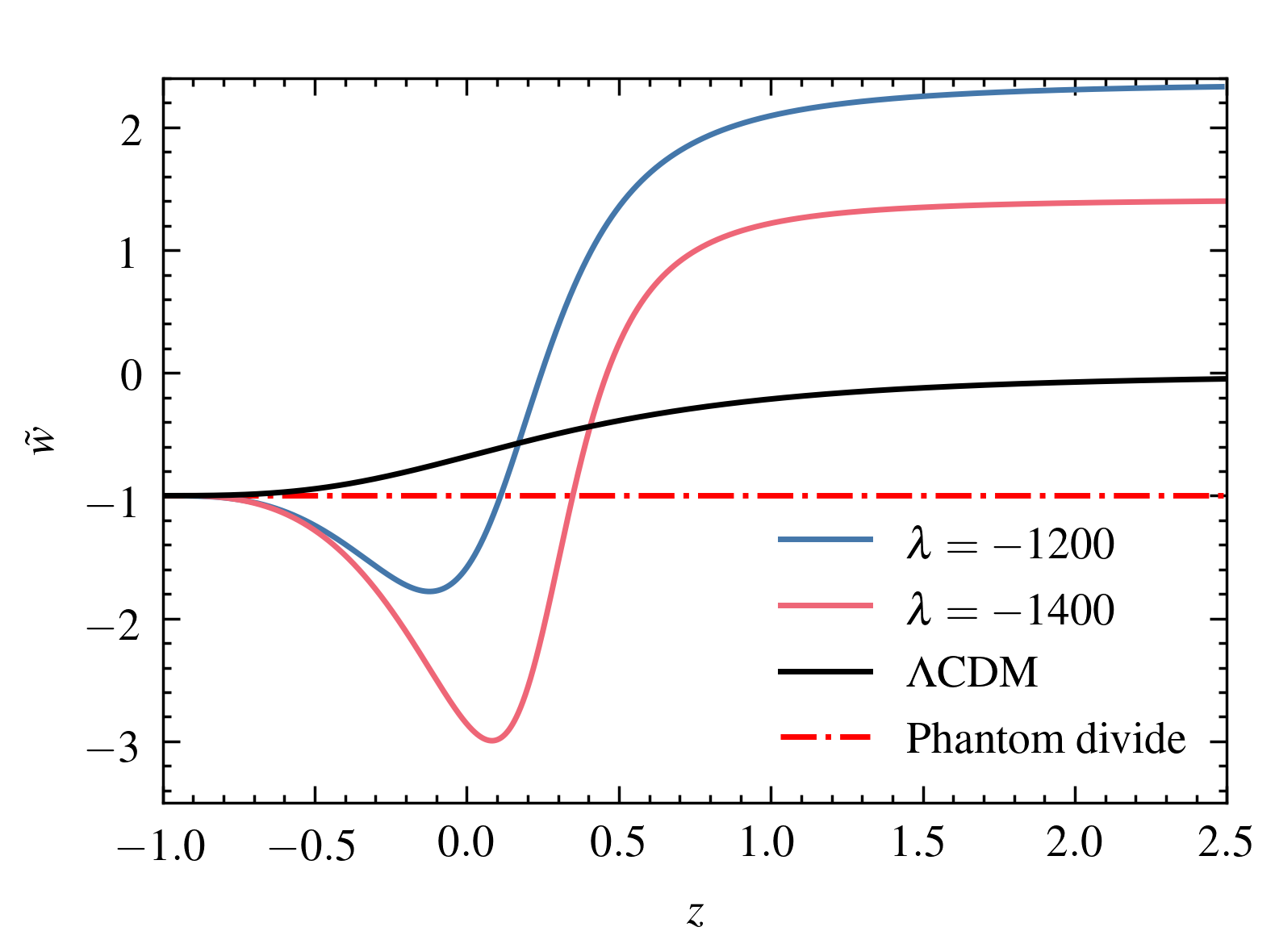}
	\caption{Total equation of state ($\tilde{w}$) as a function of redshift ($z$) for different values of $\lambda$. Phantom divide is at $\tilde{w}=-1$ (with $\Omega_{m0}=0.315$ and $H_0=67.4$ km/s/Mpc \cite{aghanim2020planck}).}
	\label{fig:new_tEoSMA}
\end{figure}

Moreover, we can evaluate the behaviour of the total equation of state parameter ($w$). For this, we first use,
\begin{equation}
w=\frac{2}{3}\left(q-\frac{1}{2}\right).
\label{eq:TEoSRHDE}
\end{equation}
The validity of the above expression depends on whether we consider the standard Friedmann equations or not. The behaviour of $w$ is similar to $q$, as the above equation indicates a linear relationship between them. Figure (\ref{fig:ModelAw}) illustrates the behaviour of $w$ for different values of $\lambda$. Additional to the total equation of state parameter, one can also evaluate the equation of state parameter for the dark energy component using the relation $w_{_\Lambda}={p_{_{\Lambda }}}/{\rho_{_{\Lambda }}}$. In a $3+1$ dimensional spacetime, from the continuity equation for dark energy density, we have,
\begin{align}
&\dot{\rho_{_\Lambda}}=-3H(\rho_{_\Lambda}+p_{_\Lambda})\implies p_{_\Lambda}=\frac{\dot{\rho}_{_\Lambda}}{-3H}-\rho_{_{\Lambda }}
\nonumber\\&\implies w_{_\Lambda}=-1+\frac{\dot{\rho}_{_\Lambda}}{-3\rho_{_{\Lambda }}H}
\end{align}
One can see from the behaviour of $w_{_\Lambda}$ in figure (\ref{fig:ModelAw}) that the dark energy equation of state crosses the phantom divide and approach the phantom divide as $z\rightarrow-1$. It is interesting to note that $w$ never crosses the phantom divide ($w=-1$) while $w_{_\Lambda}$ does. This is due to the linear relation between $q$ and $w$, as $q$ never crosses the phantom divide, nor does $w$. This feature is the consequence of assuming the validity of standard Friedmann equations. On the other hand, for calculating $w_{_\Lambda}$, we used the continuity equation instead of the second Friedmann equation. This is why we could see the Phantom behaviour of the dark energy density. 
This is the first logical inconsistency we encounter in all standard HDE models. To clarify, let us calculate the total equation of state parameter using the continuity relation. When we consider the exact relation $w=p_{_\Lambda}/(\rho_{m}+\rho_{_{\Lambda}})$, without assuming the validity of the standard Friedmann equations, we cannot arrive at the relation (\ref{eq:TEoSRHDE}). To see this, consider the expression, $p_{_\Lambda}=({\dot{\rho}_{_\Lambda}}/{(-3H)})-\rho_{_{\Lambda }}$ from the continuity relation of dark energy. Now the new expression for total equation of state parameter  ($\tilde{w}$) becomes,
\begin{equation}
\tilde{w}=\frac{{\dot{\rho}_{_\Lambda}}+3H\rho_{_{\Lambda }}}{-3H(\rho_{_{\Lambda}}+\rho_{m})}.
\end{equation}
This expression crosses the phantom divide which one can see from figure (\ref{fig:new_tEoSMA}). 
Further, the total equation of state parameter gives a higher value in the matter-dominated era, where we expect it to be zero. Thus we end up in a situation where we get two different expressions for the same total equation of state parameter. Thus, it is clear that there are inconsistencies in assuming the validity of standard Friedmann equations and the continuity equation simultaneously.

Now, why does the total equation of state parameter goes above zero in figure (\ref{fig:new_tEoSMA})? This must be because of the nature of dark energy itself. If the matter is not the dominant component in the early (not as early as the radiation dominant period) epoch, then the behaviour of dark energy affects the total equation of state. To see this, we study the behaviour of the dark energy density parameter given as
\begin{equation}
\Omega_{_\Lambda}=\frac{8\pi\rho_{_{\Lambda }}}{3H^2}=\frac{H^2(1-\Omega_{m0})}{H_0^2\log\left(1+\frac{\lambda\pi}{H_0^2}\right)}\log\left(1+\frac{\lambda\pi}{H^2}\right).
\end{equation}

\begin{figure}[h]
	\centering
	\includegraphics[width=0.45\textwidth]{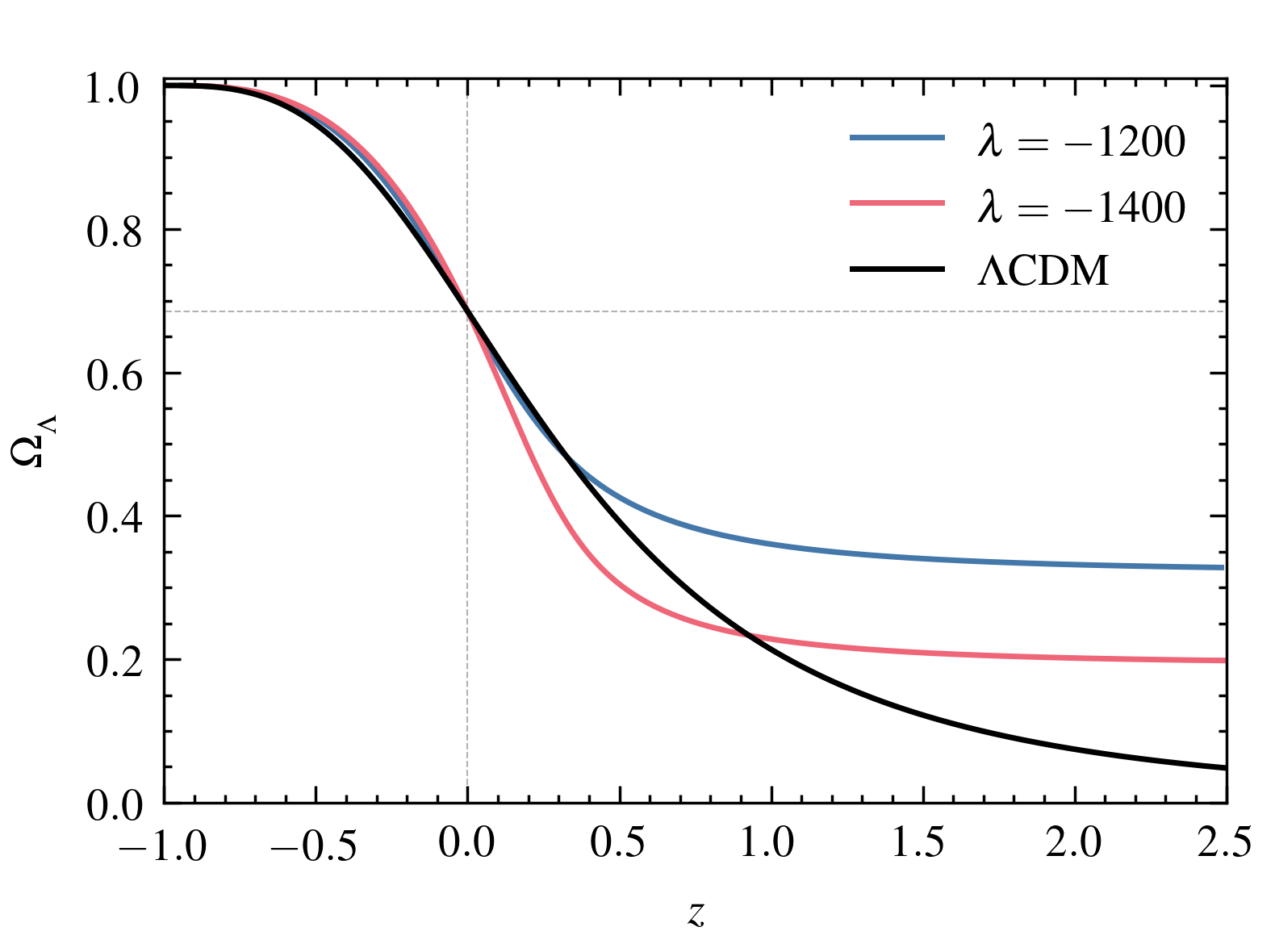}
	\caption{Dark energy density parameter ($\Omega_{_\Lambda}$) as a function of redshift ($z$) for different values of $\lambda$ (with $\Omega_{m0}=0.315$ and $H_0=67.4$ km/s/Mpc \cite{aghanim2020planck}).}
	\label{fig:6RHDEMA}
\end{figure}

Unlike the $\Lambda$CDM model, here, the value of $\Omega_{_\Lambda}$ do not tend to zero for high redshift. Instead, it settles at a non-zero value (See figure~(\ref{fig:6RHDEMA})).
From the above expression, for $z\rightarrow\infty$, we have,
\begin{equation}
\Omega_{_\Lambda}=\frac{\pi \lambda \left(1 - \Omega_{m0}\right)}{H_{0}^{2} \log{\left(1 + \frac{\pi \lambda}{H_{0}^{2}} \right)}}
\end{equation}
Thus for a fixed value of $\lambda$, $\Omega_{_\Lambda}$ remains a constant for large $z$. This effectively gives the wrong total equation of state, when we consider the relation $w=p_{_\Lambda}/(\rho_{m}+\rho_{_{\Lambda}})$. Because in the matter-dominated era, total pressure must be close to zero, and matter density must dominate the total energy density. Thus, even if $\tilde{w}$ goes above zero, it should at least come down to zero for large values of $z$. Here, since $\Omega_{_\Lambda}$ never approaches zero, it contributes to the total equation of state parameter. Thus, simultaneously considering the validity of Friedmann equations and the validity of continuity equations for HDE models may lead to counter-intuitive results. All of this demands us to reconsider the standard HDE approach.

So far, we did not bother about the possible values of $\lambda$. In the equation (\ref{eq:RHDE_sDE}), the dependency on $\lambda$ comes from the factor ${\log\left(1+\frac{\lambda\pi}{H^2}\right)}/{\log\left(1+\frac{\lambda\pi}{H_0^2}\right)}$. Now, by the conventional definition of R\'enyi entropy, the value of $\lambda$ must lie between $1$ and $-\infty$ to be Schur concave. A stronger concavity condition demands $\lambda$ to lie between $1$ and zero. However, positive values, such that $\lambda>1$ is in principle allowed because of the skew symmetry of R\'enyi entropy about $\lambda=1/2$. An interesting situation will arise in our case when $\lambda\pi/H^2<-1$. In such cases, the dark energy density defined using equation (\ref{eq:RHDE_sDE}) will be complex. Thus, the value of $\lambda$ cannot be arbitrarily negative. Thus, mathematically, the allowed values for $\lambda$ must be between $+\infty$ to $-H^2/\pi$. However, in Moradpour et al.'s approach, this problem never arises, as there is no log function present in their expression (See equation (14) in \cite{Moradpour2018}). Article \cite{refId0} also demands a very tiny positive value for $\lambda$ in the context of holographic equipartition due to Padmanabhan \cite{padmanabhan2012emergence}.

\subsection{Thermodynamics near Phantom divide}

In the model discussed above, for certain values of $\lambda$, the dark energy equation of state appears to cross the Phantom divide and then reaches the final de Sitter state in the asymptotic future. Since Phantom models are not ruled out by current observations, these results needs some extra attentions. Since our primary motive is to have a consistent thermodynamic picture, let us explore the thermodynamics at the Phantom divide by asking ourself certain questions.

\subsubsection{What happens to the temperature at $w=-1$?}

In this article we adapt two definitions for horizon temperature, depending whether we use the Clausius relation or unified first law. While using the Clausius relation ($-dE=TdS$) we have, $T=1/\left(2\pi{\tilde{r}_{\!\!_A}}\right)$. Thus, by definition, the temperature does not become zero or negative throughout the entire course of evolution. This temperature also goes by the name ``\textit{Cai and Kim temperature}'', as it was introduced in \cite{Cai_2005}.

When the model is at the Phantom divide, i.e. $w=-1$, we have $p:=w\rho=-\rho$ and we get $-dE=0$. This result implies that either $T$ or $dS$ has to be zero. Since by definition $T$ is always positive-non-zero (except for the limit ${\tilde{r}_{\!\!_A}}\rightarrow\infty$), the only choice is to assume that $dS=0$. Thus, the third law of thermodynamics is perfectly respected and the second law approaches its limit. When the model crosses the Phantom divide (i.e. $w<-1$), we will have a situation in which the entropy decreases. This situation is because $dE$ is positive for the outward flux and negative for the inward flux by convention \cite{PhysRevD.92.024001}. With $dE$ as a positive quantity, $dS$ becomes negative and the entropy decreases. Hence, the second law of thermodynamics is not respected, when $w<-1$.

If we were to start from the unified first law, which states that $dE=TdS+WdV$, with $W=(\rho-p)/2$, the temperature from the induced flat FLRW metric is, 
\begin{equation}
T=\frac{-1}{2\pi{\tilde{r}_{\!\!_A}}}\left(1-\frac{\dot{{\tilde{r}_{\!\!_A}}}}{2H{\tilde{r}_{\!\!_A}}}\right)=\frac{-1}{2\pi{\tilde{r}_{\!\!_A}}}\left(1-\frac{\dot{H}}{2H^2}\right)
\end{equation}
In the standard HDE adapted above, we assumed the validity of the standard Friedmann equations. Thus, one can rewrite the above temperature in terms of $w$ as, 
\begin{equation}
T=\frac{-1}{2\pi{\tilde{r}_{\!\!_A}}}\left(\frac{1-3w}{4}\right)
\label{eq:tempwithw}
\end{equation}
Then, at the phantom divide ($w=-1$) we have $T=-1/(2\pi{\tilde{r}_{\!\!_A}})$ and $W=\rho$. Now the unified first law reduces to, $dE=TdS+\rho dV$. Since, $E=\rho V$ (the Misner-Sharp energy) and $dE=Vd\rho+\rho dV$, we have, $TdS=Vd\rho$. In the dark energy dominated epoch (i.e. $\rho\approx\rho_{\Lambda}$) the density $\rho$ is a constant when $w=-1$ which implies $d\rho=0$. Thus we have $TdS=0$. Based on the previous arguments, temperature, by definition, remains non-zero while entropy is at its extremum. Further, if $w$ becomes less than unity, entropy decreases and the second law no longer holds. However, unlike some Phantom models with $w<-1$ for $z\rightarrow-1$, the above model shows a final de Sitter space. Thus all the laws of thermodynamics are respected in the far future. This might indicate a deviation from the standard equilibrium thermodynamics \cite{PhysRevLett.96.121301} on the course of cosmic evolution.

Another interesting observation would be for the radiation dominant era. Here, the equation of state will be $w=1/3$, and the Cai and Kim temperature posses no problem, as it is always well defined. However, the temperature defined using the induced flat FLRW metric, equation (\ref{eq:tempwithw}), becomes zero. Thus, there is a sign flip for the temperature at $w=1/3$. At this point (also known as the Hayward zero temperature divide), ${\tilde{r}_{\!\!_A}}$ is a null surface that divides the spacelike and timeline regions, and it does not respect the third law of thermodynamics \cite{PhysRevD.92.024001}. The systematic study of radiation dominated epoch in the concerned HDE is kept for elsewhere.

\subsubsection{Effects of interacting dark energy}

An important caveat in the above discussion was the absence of matter - dark energy interaction. So far there are no observational evidence which disprove the existence of non-standard interaction at cosmic scale.  In fact this gives us the freedom to have scalar field, vector fields etc. as our dark energy candidates.  Further, as  the nature of interaction is unknown, nothing stops us from considering linear as well as non-linear interaction between dark matter and dark energy, as long as we can explain the late time acceleration and structure formations \cite{Cabral_2009,Farrar_2004}. 

Here, instead of the standard conservation equations, one has, $\dot{\rho_i}+3H(1+w_i)\rho_i=Q$, where $Q$ is the interaction term, which can be either linear or non-linear functions of densities along with unknown constants. Once there is interactions, it is possible to set $w=-1$, by which the model reduces to $\Lambda$CDM when $Q=0$ (see \cite{doi:10.1142/S0218271822501073} for example). Thus the model never crosses the Phantom divide by definition. One can also rule out certain interactions, as it cannot explain the power spectrum \cite{PhysRevD.106.043527}. One way to test Phantom models is to investigate the density perturbations and check for the stability. This depends on the form of interaction and is subjected to its compatibility with energy conditions \cite{Jimenez_2003}. Reviews by Wand et. al \cite{Wang_2016} and Bolotin et. al \cite{doi:10.1142/S0218271815300074} gives a comprehensive treatment of the subject. A detailed analysis including various interaction will not be in the rest of the article.

\section{A holographic dark energy from entropic function}

So far, the holographic principle and the laws of thermodynamics appears far from each other, and there are many inconsistencies in the standard HDE approach. In this section we propose a novel way to model a dark energy density from the entropy correction factors appearing in the modified Friedmann equations. Our approach will respect both holographic bound and the laws of thermodynamics.

Clearly, we get additional terms in the modified Friedmann equations when we consider entropies other than the Bekenstein-Hawking entropy. One can interpret these additional factors as the source of holographic dark energy. Very recently, authors in references \cite{PhysRevD.105.044042, NOJIRI2022137189, NOJIRI2022136844} adopted a similar approach to explain the cosmic evolution. They equated the entropic correction factor with the standard HDE density and redefined the IR cutoff. They illustrate the possibility of explaining the cosmic evolution with and without the integration constant, which is essentially the cosmological constant. Here, instead of equating the extra entropic function with the standard HDE density, we assume its proportionality with the CKN bound to define a new HDE density. In the process, we drop the integration constant, which one can recover as a limiting case. 

From the laws of thermodynamics (both the Clausius relation and the unified first law), we originally had,
\begin{equation}
\rho=\frac{-1}{2\pi\Omega_n}\int\frac{1}{\tilde{r}_{\!\!_A}^{n+1}}dS.
\end{equation}
This is essentially the first Friedmann equation. Upon integrating the right-hand side, we get the standard $\sim H^2$ term, additional factors, and an integration constant. In the $\Lambda$CDM model, this integration constant is responsible for the late time acceleration. Thus, to recover the factor responsible for the cosmic acceleration, we must remove the standard Bekenstein-Hawking factor from the integral. In a more general setting, we argue that one must remove the contribution from the standard area law to extract the explicit contribution from the entropic correction factor. Using this notion, we propose that the dark energy must be of the form,
\begin{align}
\rho_{_{\Lambda}}\propto\left(\frac{-1}{2\pi\Omega_n}\int\frac{1}{{\tilde{r}_{\!\!_A}}^{n+1}}dS\right)-\frac{n(n-1)}{16\pi{\tilde{r}_{\!\!_A}}^2}.
\label{eq:new_bound1}
\end{align}
Here, ${n(n-1)}/{(16\pi{\tilde{r}_{\!\!_A}}^2)}$ removes the terms coming from the Bekenstein-Hawking entropy. This proposal is dimensionally consistent with the holographic dark energy and gives the cosmological constant in the $\Lambda$CDM model when $S=\pi\tilde{r}_{\!\!_A}^{2}$. Besides, the holographic principle is a bound rather than a functional form. Since we remove the contribution from the Bekenstein-Hawking term, the effective functional form in equation (\ref{eq:new_bound1}) is well under the allowed CKN bound. Thus respecting the holographic principle and thermodynamics simultaneously. Other HDE models also show $\Lambda$CDM-like behaviour; however, such models assumed $w=-1$ from the beginning, which is a trademark feature of $\Lambda$CDM \cite{Dheepika2022a, doi:10.1142/S0218271822501073}. Now, using the notion of holography, we define the dark energy density as,
{\begin{align}
	\rho_{_{\Lambda}}=C^2\left[\left(\frac{-1}{2\pi\Omega_n}\int\frac{1}{{\tilde{r}_{\!\!_A}}^{n+1}}dS\right)-\frac{n(n-1)}{16\pi{\tilde{r}_{\!\!_A}}^2}\right].
	\label{eq:NEWHDE}
	\end{align}}
For entropies other than Bekenstein-Hawking area law, we get additional factors, and it is then possible to set the integration constant to zero by defining $C^2$ using the conservation relation,
\begin{equation}
\Omega_{_\Lambda}+\Omega_{m}=1,
\end{equation}
for a flat universe ($k=0$). Here, $\Omega_m = {8\pi\rho_m}/({3H^2})$ and $\Omega_{_\Lambda} = {8\pi\rho_{_\Lambda}}/({3H^2})$ are the dimensionless density parameters. It is useless if our HDE model cannot explain cosmic evolution. Hence, in the next section, we discuss the cosmic evolution by using the R\'enyi entropy as the choice.

\section{New holographic dark energy with R\'{e}nyi entropy}

Out of all HDE densities discussed earlier, R\'enyi entropy was exceptional, and the standard approach to R\'enyi HDE is not satisfactory. Here, we use our definition of dark energy density, given in equation (\ref{eq:NEWHDE}), introduced in the previous section to define an HDE density. Here, we consider the validity of the continuity equation and the first law of thermodynamics. From equation (\ref{eq:NEWHDE}), we have the definition of new holographic dark energy as
\begin{equation}
\rho_{_{\Lambda }}=C^2\left[\left(\frac{-1}{2\pi\Omega_n}\int\frac{1}{{\tilde{r}_{\!\!_A}}^{n+1}}dS\right)-\frac{n(n-1)}{16\pi{\tilde{r}_{\!\!_A}}^2}\right].
\end{equation}
Considering a flat $3+1$ dimensional FLRW universe with R\'enyi entropy as the choice, we get,
\begin{equation}
\rho_{_{\Lambda }}=-\frac{3 C^{2} }{8 \pi^{2}}\left[ \pi^2\lambda\log\left(1+\frac{1}{\pi\lambda{\tilde{r}_{\!\!_A}}^2}\right)\right]+\tilde{\Lambda}
\end{equation}
With the integration constant $ \tilde{\Lambda} $, we will get acceleration, where the extra factor acts like a small correction. Here, we set  $\tilde{\Lambda}=0$ and explain the cosmic evolution from the correction factor alone. To do so, we use the relation $\Omega_{_\Lambda}+\Omega_{m}=1$, and with $\tilde{r}_{\!\!_A}=1/H$, the expression for $C$ becomes,
\begin{equation}
C=\sqrt{\frac{H_{0}^{2} \left(1 - \Omega_{m0}\right)}{\pi \lambda \log{\left(\frac{\pi \lambda}{H_{0}^{2} + \pi \lambda} \right)}}}
\end{equation}
Here, $\Omega_{m0}$ and $H_0$ are the present value of the matter density parameter and the Hubble parameter, respectively. Thus the expression connecting Hubble parameter, energy densities, and redshift becomes,
\begin{equation}
h(z)^{2} = \Omega_{m0} \left(z + 1\right)^{3} + \frac{\left(1 - \Omega_{m0}\right) \log\left(\frac{\pi\lambda}{H^2+\pi\lambda}\right)}{\log\left(\frac{\pi\lambda}{H_{0}^2+\pi\lambda}\right)}
\end{equation}
Interestingly, in the limit $\lambda\rightarrow0$, we get, $h(z)^{2} = \Omega_{m0} \left(z + 1\right)^{3} + \left(1 - \Omega_{m0}\right) $,
which is precisely the $\Lambda$CDM model. Once again, it is not easy to obtain an analytical solution for $\lambda\ne0$; hence we rely on numerical techniques. 
\begin{figure}[h]
	\centering
	\includegraphics[width=0.45\textwidth]{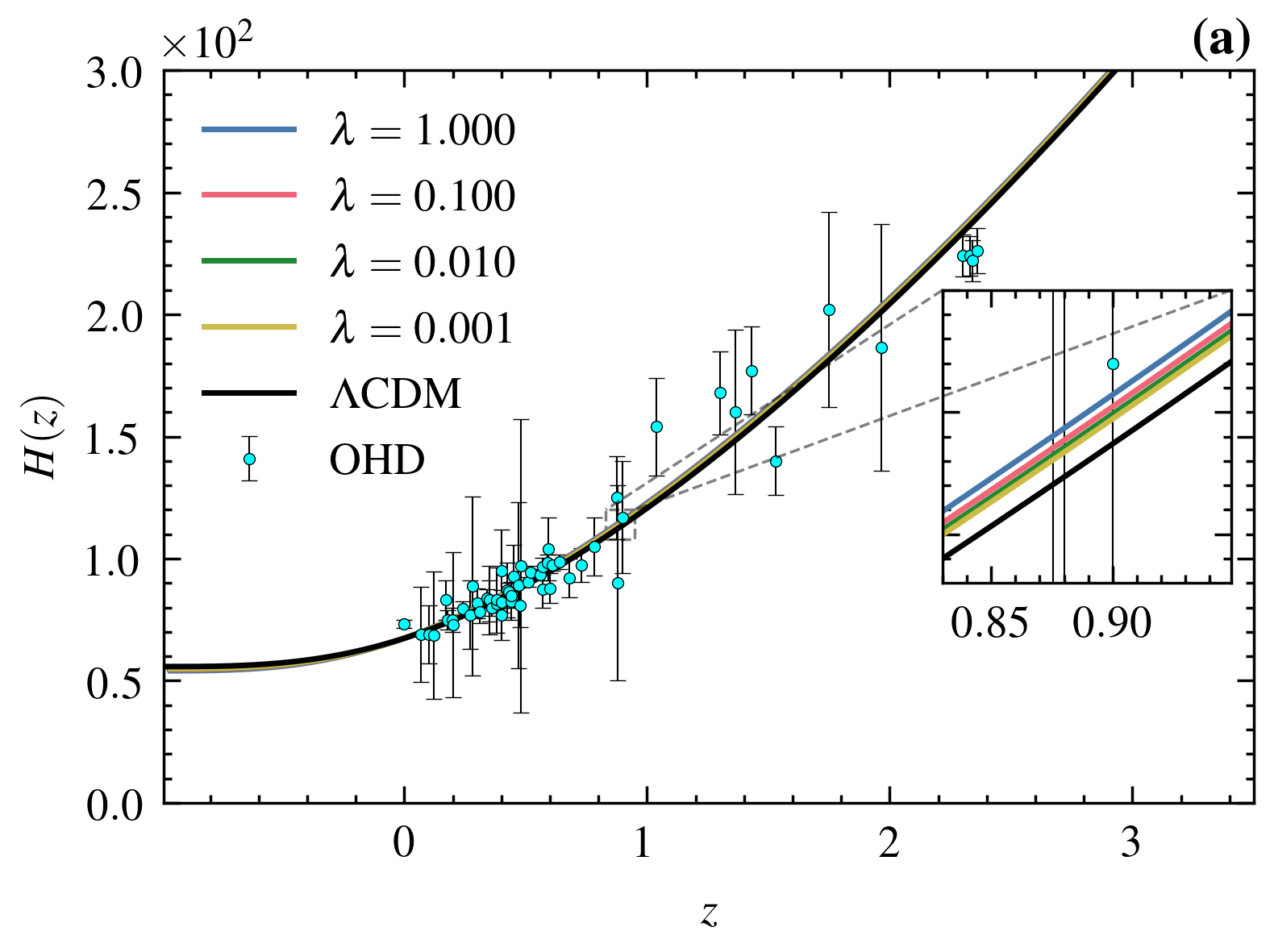}
	\includegraphics[width=0.45\textwidth]{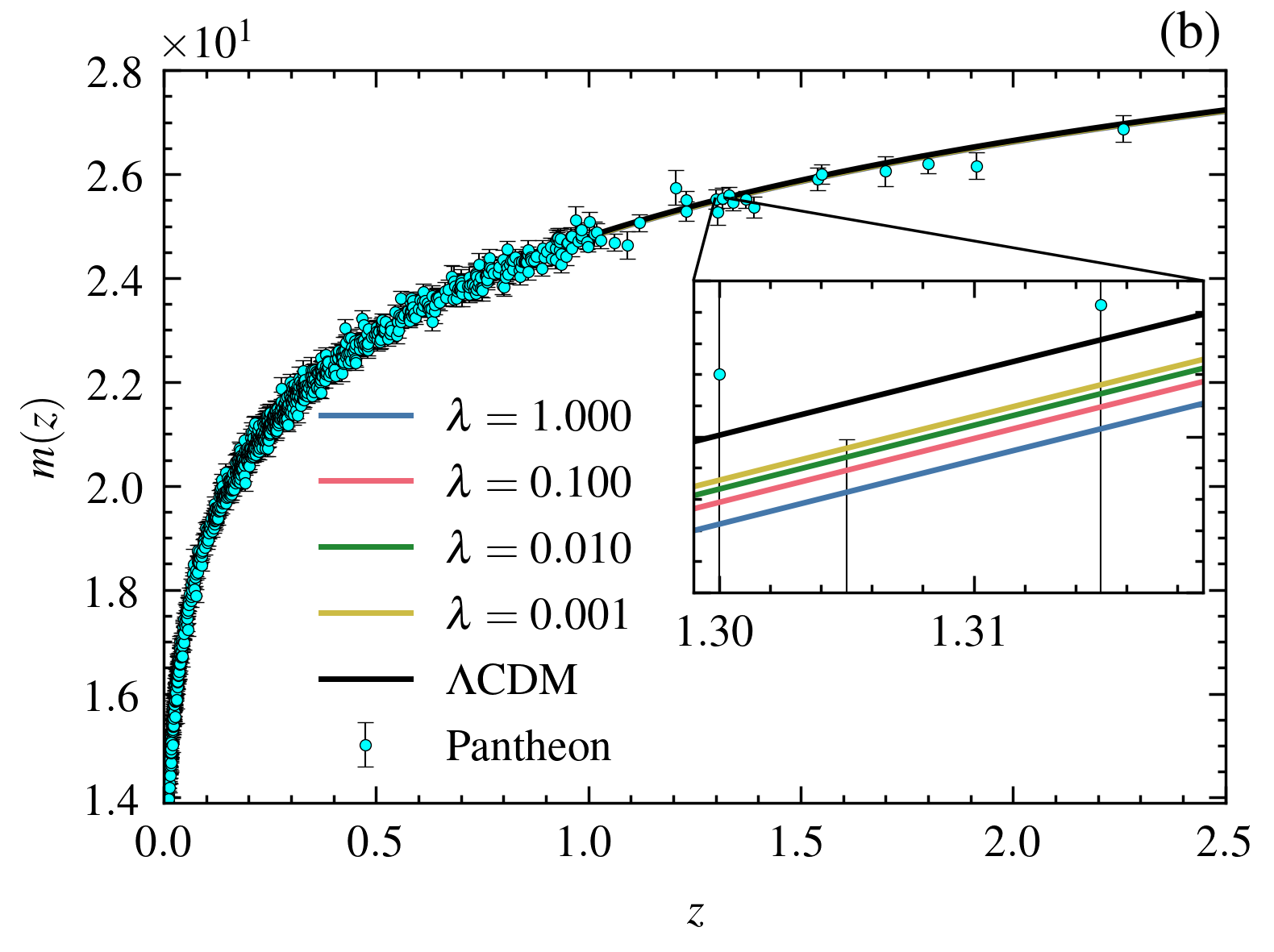}
	\caption{(a) Hubble parameter ($H(z)$) and (b) apparent magnitude ($m(z)$) of Type-1a supernovae as a function of redshift ($z$) for different values of $\lambda$. Here we took $M=-19.366$ as the nuisance parameter \cite{10.1093/mnras/stab3773} to estimate the apparent magnitude (with $\Omega_{m0}=0.315$ and $H_0=67.4$ km/s/Mpc \cite{aghanim2020planck}). Observational Hubble Data from \cite{https://doi.org/10.1002/asna.20220003} and Pantheon data from \cite{Scolnic_2018}. }
	\label{fig:7}
\end{figure}
Figure (\ref{fig:7}) shows the behaviour of the Hubble parameter and the apparent magnitude as a function of redshift. Here, the expression for apparent magnitude is, $m(z)=5\log_{10}\left[c(1+z)\int_{0}^{z}\frac{dz'}{H(z')}\right]+25+M$, where $M$ is the nuisance parameter. Clearly, for a very small value of $\lambda$, the model behaves very close to the $\Lambda$CDM model. As $\lambda$ increases, the model deviates from the $\Lambda$CDM, and for a very high value of $\lambda$, it will approach a final de Sitter universe.

\paragraph{\textbf{Value of `$\lambda$'}:--} A complete data analysis using the available cosmological observation is a rigorous work that needs more attention beyond the scope and length of the current manuscript. However, we expect $\lambda$ to be very small based on the following reasons.

By the definition of R\'enyi entropy, $\lambda$ must lie between $1$ and $-\infty$ to be Schur concave. An even stronger concavity condition from quantum information theory demands $\lambda$ to be between 1 and 0 \cite{RevModPhys.81.865}. Positive values greater than one are mathematically allowed due to the skew symmetry of R\'enyi entropy. Although there is no physical motivation to assume so. Even if we relax the stronger concavity condition and assume negative values, for $\lambda\pi/H^2<-1$, the R\'enyi entropy becomes complex-valued. Such entropies are of interest in information theory  \cite{Nalewajski2016,doi:10.1080/03610926.2022.2094963}, and it will be interesting to investigate them in the context of cosmology and black hole physics. In summary, mathematically the uniform prior range of $\lambda$ can be between $+\infty$ to $-H^2/\pi$ if we restrict to real valued entropy, and physically, based on quantum information theory, the uniform prior range with stronger concavity is between 0 and 1. A freehanded range would be $-H^2/\pi$ to 1, thus making $\lambda\rightarrow0$ a special case rather than a limit. Since such a lower bound is time dependent, we did not used this as we already assume $\lambda$ as an unknown constant. A time dependent non-extensive parameter is interesting to investigate and will be done elsewhere. N. Komatsu \cite{refId0} illustrated a behaviour similar to ours with R\'enyi entropy in the context of Padmanabhan's law of emergence \cite{padmanabhan2012emergence}. Although different from our approach, they also preferred a very small positive value for $\lambda$.

\section{Late time cosmological behaviour}
In this section we study various cosmological parameters to understand the late time behaviour of the proposed dark energy model. We will focus on the behaviour of the deceleration parameter ($q$), dark energy equation of state parameter ($w_{_\Lambda}$),  dark energy density parameter ($\Omega_{_\Lambda}$), sound speed squared ($v_s^2$), and the statefinder parameters $\{r,s\}$.

Now, using the expression $q=-1+\frac{1+z}{H(z)}\left(\frac{dH(z)}{dz}\right)$,
we can evaluate the deceleration parameter. 
\begin{figure}[h]
	\centering
	\includegraphics[width=0.45\textwidth]{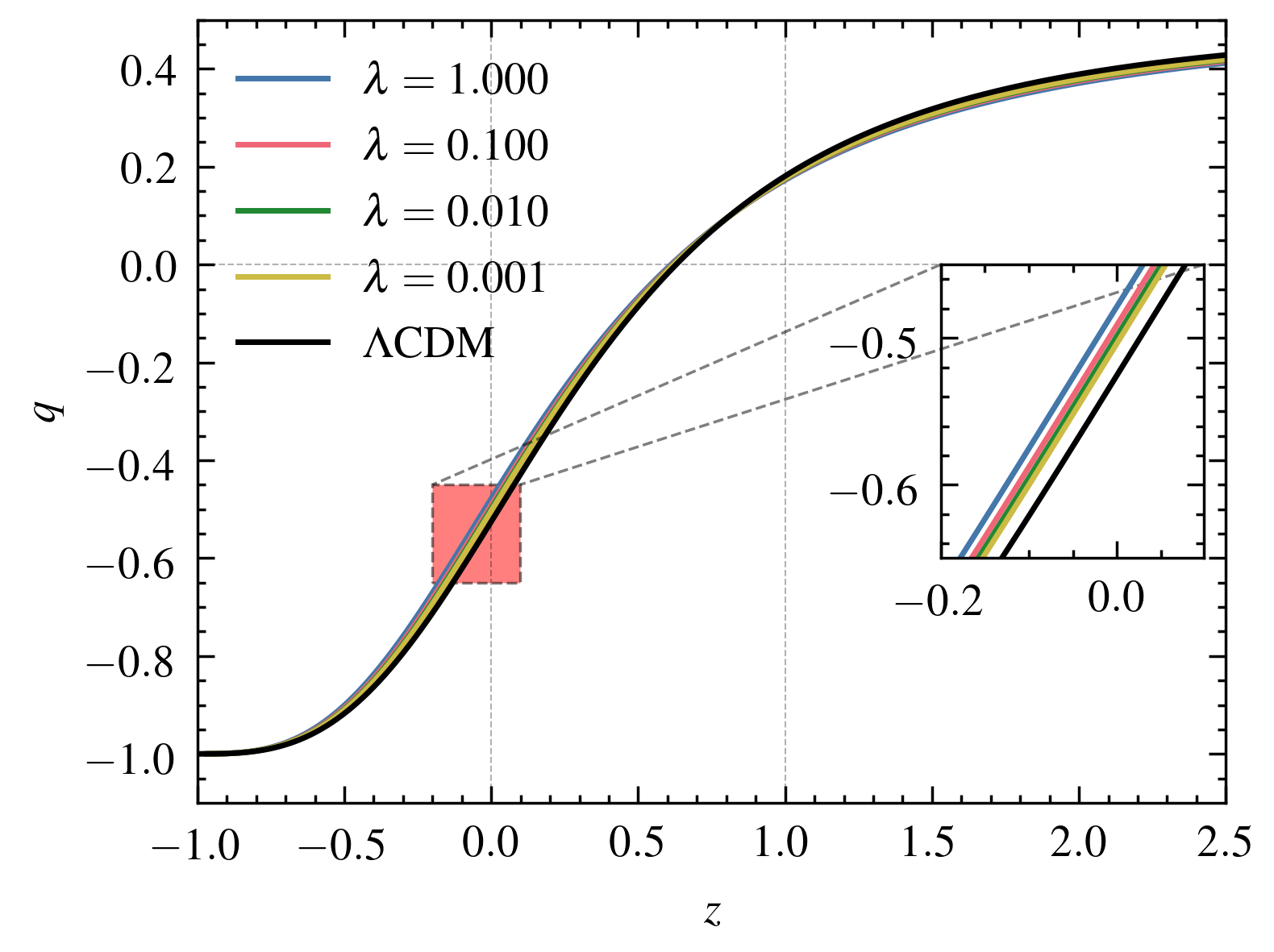}
	\includegraphics[width=0.45\textwidth]{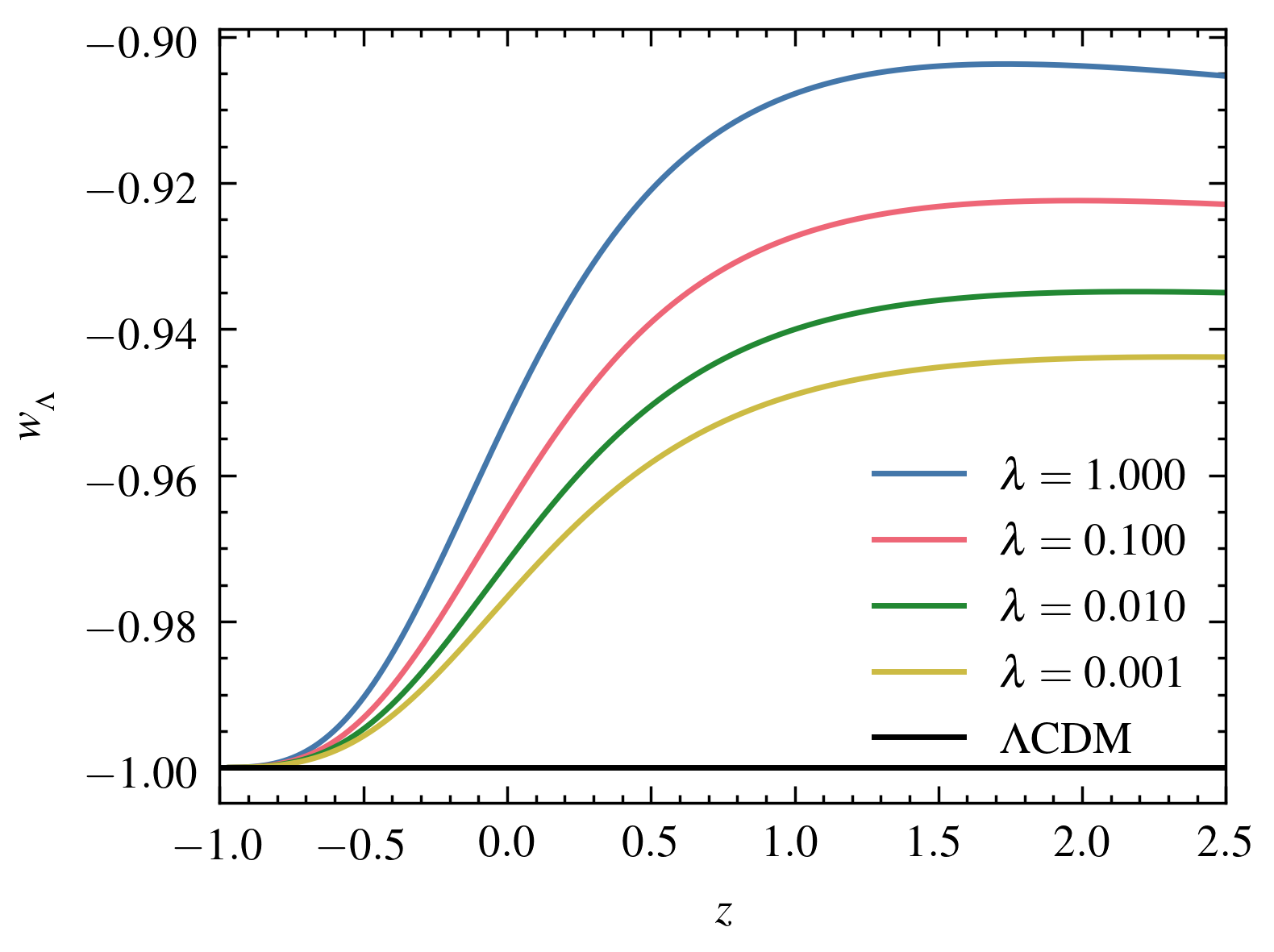}
	\caption{Deceleration parameter ($q$) and dark energy equation of state parameter ($w_{_\Lambda}$) as a function of redshift ($z$) for different values of $\lambda$ (with $\Omega_{m0}=0.315$ and $H_0=67.4$ km/s/Mpc \cite{aghanim2020planck}).}
	\label{fig:8}
\end{figure}
From the figure (\ref{fig:8}), the model shows late time acceleration around redshift $z\sim 0.5$ and tends to a final de Sitter phase for a very small positive value of $\lambda$. For very large values of $\lambda$, the model could re-enter the decelerated phase in the future. Thus, the model prefers a very small value of $\lambda$, indicating evolution close to the $\Lambda$CDM model.

\begin{figure}[h!]
	\centering
	\includegraphics[width=0.45\textwidth]{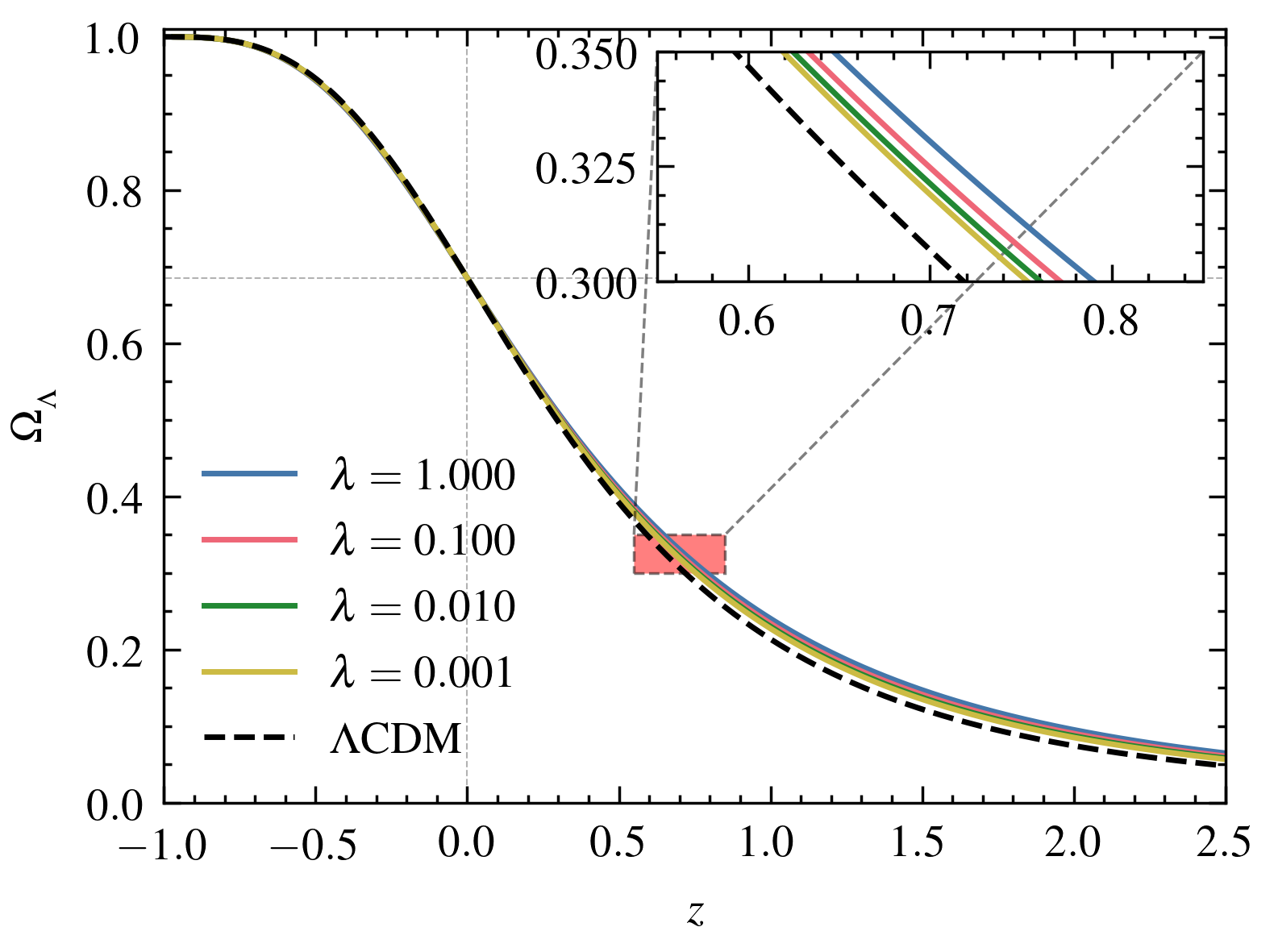}
	\caption{Dark energy density parameter ($\Omega_{_\Lambda}$) as a function of redshift ($z$) for different values of $\lambda$ (with $\Omega_{m0}=0.315$ and $H_0=67.4$ km/s/Mpc \cite{aghanim2020planck}).}
	\label{fig:9}
\end{figure}

Using the standard definition $w_{_\Lambda}=p_{_\Lambda}/\rho_{_{\Lambda }}$, where $p_{_\Lambda}=-(\dot{\rho}_{\Lambda}/(3H))-\rho_{\Lambda}$, we numerically evaluate and plot the evolution of dark energy equation of state parameter as a function of $z$ in figure (\ref{fig:8}). Clearly, for $\lambda\le1$, the model never crosses the phantom divide. Unlike the previous case of the standard R\'enyi HDE model, the dark energy density parameter $\Omega_{_\Lambda}=8\pi\rho_{_{\Lambda }}/(3H^2)$ reduces to zero for large $z$, indicating a proper matter dominated early phase. Here we have,
\begin{equation}
\Omega_{_\Lambda}=\frac{H_0^2\left(1 - \Omega_{m0}\right) \log\left(\frac{\pi\lambda}{H^2+\pi\lambda}\right)}{H^2\log\left(\frac{\pi\lambda}{H_{0}^2+\pi\lambda}\right)}
\end{equation}
For $\lambda\rightarrow0$, the above expression reduces to $H_0^2(1-\Omega_{m0})/H^2$, which is the $\Lambda$CDM behaviour. Further, it is evident from the above expression and the figure (\ref{fig:9}) that $\Omega_{_\Lambda}$ goes to zero, rather than settling in a non-zero value for large $z$.  

\begin{figure}[t]
	\centering
	\includegraphics[width=0.45\textwidth]{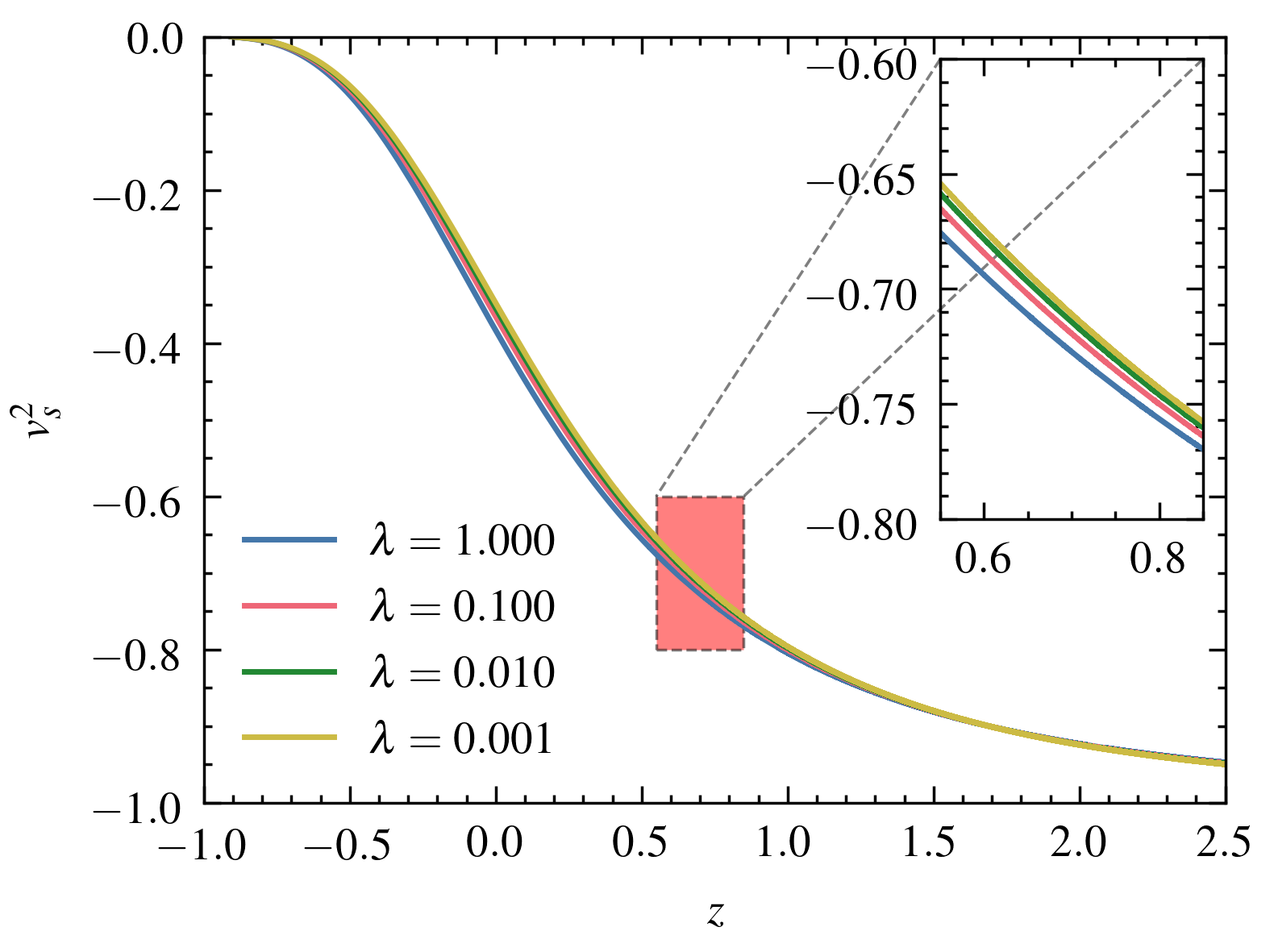}
	\caption{Sound speed squared ($v_s^2$) as a function of redshift ($z$) for different values of $\lambda$ (with $\Omega_{m0}=0.315$ and $H_0=67.4$ km/s/Mpc \cite{aghanim2020planck}).}
	\label{fig:10}
\end{figure}

Now we evaluate the stability of the model by calculating the sound speed square ($v_s^2$) given as,
\begin{equation}
v_s^2=\frac{dp_{_\Lambda}}{d\rho_{_{\Lambda }}}=w_{_\Lambda}+\rho_{_{\Lambda }}\frac{dw_{_\Lambda}}{d\rho_{_{\Lambda }}}.
\end{equation}
For a stable model, the value of $v_s^2$ lie between the region zero and 1. Values beyond this region indicate instabilities of various kinds. We can see from the figure (\ref{fig:10}) that the model goes from an unstable state to a final stable de Sitter limit. 

\begin{figure}[h]
	\centering
	\includegraphics[width=0.45\textwidth]{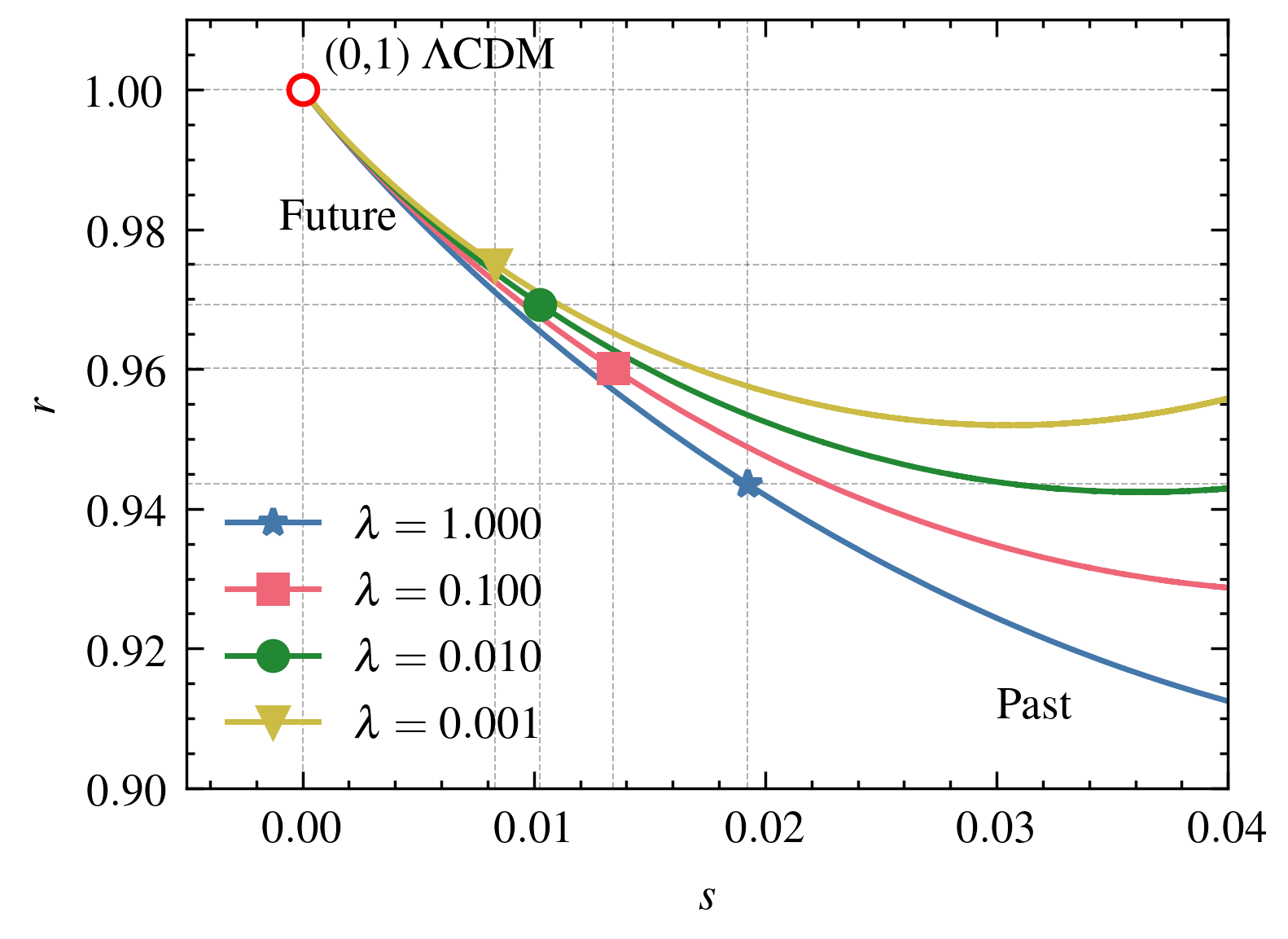}
	\caption{Statefinder diagnostic for different values of $\lambda$. The sectors divided by dotted lines about the markers represents the past and the future  (with $\Omega_{m0}=0.315$ and $H_0=67.4$ km/s/Mpc \cite{aghanim2020planck}). The markers represents the present time for different $\lambda$.}
	\label{fig:11}
\end{figure}
To distinguish our model from the $\Lambda$CDM model, we use the statefinder diagnostic. This geometric diagnostic tool has two parameters $\{r,s\}$ defined as,
\begin{align}
r=\frac{\dddot{a}}{aH^3}\text{ and }s=\frac{r-1}{3\left(q-\frac{1}{2}\right)}.
\end{align}
Here, $r$ is the jerk parameter. For the $\Lambda$CDM model, these parameters take a fixed value $\{r,s\}=\{1,0\}$, while for a varying dark energy model, it traces a parametric path depending on the evolution of the density parameters, the equation of state parameter and its time derivatives. In terms of scaled Hubble parameter $h(z)=H(z)/H_0$, the statefinder parameter $\{r,s\}$ takes the form
\begin{equation}
r= \frac{1}{2h^2}\frac{d^2h^2}{dx^2}+\frac{3}{2h^2}\frac{dh^2}{dx}+1
\text{ and }s = -\frac{\frac{1}{2h^2}\frac{d^2h^2}{dx^2}+\frac{3}{2h^2}\frac{dh^2}{dx}}{\frac{3}{2h^2}\frac{dH^2}{dx}+\frac{9}{2}}.
\end{equation}
Where, $x=\ln(a)$ and $a=1/(1+z)$ is the scale factor. From the analysis, our model is distinguishable from the $\Lambda$CDM model. Figure (\ref{fig:11}) shows the parametric $\{r,s\}$ plot for different values of $\lambda$. For $\lambda\neq0$, the model lies in the quintessence region, where $r<1$ and $s>0$ and goes to $\Lambda$CDM in the future.

To summarise our numerical analysis, for $\lambda\in[0,1]$, the model parameters are well within the observational bounds and very close to the concordance $\Lambda$CDM model estimates. Model independent analysis of cosmological data, such as ``Cosmography'' \cite{dunsby2016theory,Visser2005} also support our estimates \cite{doi:10.1142/S0218271819300167,10.1093/mnras/sty422,PhysRevD.86.123516,PhysRevD.89.103506}. The following table tabulates various estimates for different values of $\lambda$ based on the model discussed above. Here we used $\Omega_{m0}=0.315$ and $H_0=67.4$km/s/Mpc from the Planck 2018 result \cite{aghanim2020planck}.
\begin{table}[h!]
	\centering
	\begin{tabular}{ccccc}
		\hline
	$\lambda$	& $q_0$ & $z_t$  &$w_{_{\Lambda0}}$  & $r$ \\
		\hline
		1.000& -0.4784 & 0.6118  &-0.9522  & 0.9435  \\
	
		0.100& -0.4911 &0.6166  & -0.9646 &  0.9601 \\
	
		0.010& -0.4985 &0.6196  &-0.9719  & 0.9692   \\
		
		0.001& -0.5035 & 0.6116 &-0.9767  & 0.9750  \\
	
		0 & -0.5257 & 0.6256 & -1 & 1  \\
		\hline
	\end{tabular}
	\caption{Estimates of various parameters for different values of $\lambda\in[0,1]$ with $\Omega_{m0}=0.315$ and $H_0=67.4$ km/s/Mpc \cite{aghanim2020planck}. Here $\lambda$ is the model parameter, $q_0$ is the present value of deceleration parameter, $z_t$ is the transition redshift, $w_{_{\Lambda0}}$ is the present value of dark energy equation of state and $r$ is the jerk parameter (also the 1$^{st}$ statefinder parameter).}
	\label{tab:3}
\end{table}

From the above table it is clear that, for $\lambda\in[0,1]$, the deceleration parameter, $q_0\in(-0.5257,-0.4784)$, the transition redshift, $z_t\in(0.6256, 0.6118)$, dark energy equation of state, $w_{_{\Lambda0}}\in(-1, -0.9522)$ and the jerk parameter, $r\in(1,0.9435)$ for $\Omega_{m0}=0.315$ and $H_0=67.4$ km/s/Mpc.

The jerk parameter is also denoted by `$j$' instead of `$r$' in the literature \cite{Aviles2017}. However, we used `$r$' to quickly compare it with the statefinder diagnostics illustrated by V. Sahni et al. \cite{Sahni2003}. In our analysis, for the given parameter space of $\lambda\in[0,1]$, our model can explain the late time acceleration with a quintessence type dark energy, where $r$ is positive by construction. Previously, Luongo \cite{doi:10.1142/S0217732313500806} showed that a positive jerk parameter can explain the late time acceleration and departure from the $\Lambda$CDM model. Their calculations suggest that the cosmological constant can be interpreted as a limiting case of a more general dark energy model. Here in our model, the $\Lambda$CDM precisely appears as the limiting case when $\lambda\rightarrow0$, and also in the asymptotic future, when $\lambda\neq0$.

\section{Entropy evolution and the second law of thermodynamics}
Finally, we explore the thermodynamic evolution of our new model. The entropy of the universe is the sum of the entropy of the horizon and the entropy of everything inside the horizon. Since the total entropy inside the horizon is very less when compared to the horizon entropy, one may approximate the total entropy as the entropy of the horizon \cite{egan2010larger}. Also, according to the second law of thermodynamics, the entropy of an isolated system must be an increasing function of time. Here, we have R\'enyi entropy as the horizon entropy, and it reads,
\begin{equation}
S=\frac{1}{\lambda}\log\left[1+\lambda\left(\frac{\pi}{H^2}\right)\right].
\end{equation}
\begin{figure}[h]
	\centering
	\includegraphics[width=0.45\textwidth]{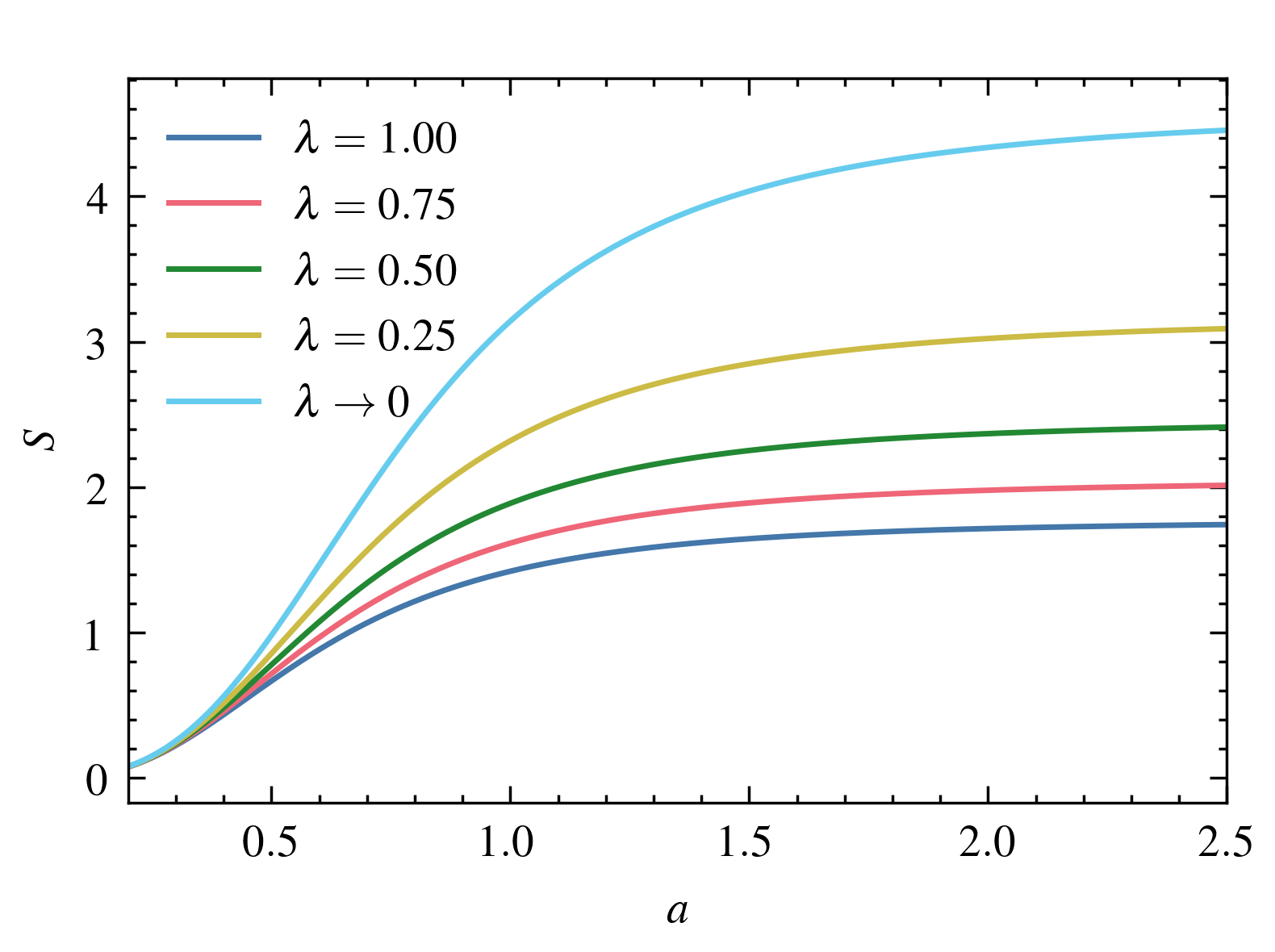}
	\caption{Horizon entropy ($S$) as a function of scale factor ($a$) for different values of $\lambda$ (with $\Omega_{m0}=0.315$ and $H_0=67.4$ km/s/Mpc \cite{aghanim2020planck}). Here, $\lambda\rightarrow0$ implies $\Lambda$CDM behaviour (in natural units).}
	\label{fig:12}
\end{figure}

\begin{figure}[h]
	\centering
	\includegraphics[width=0.45\textwidth]{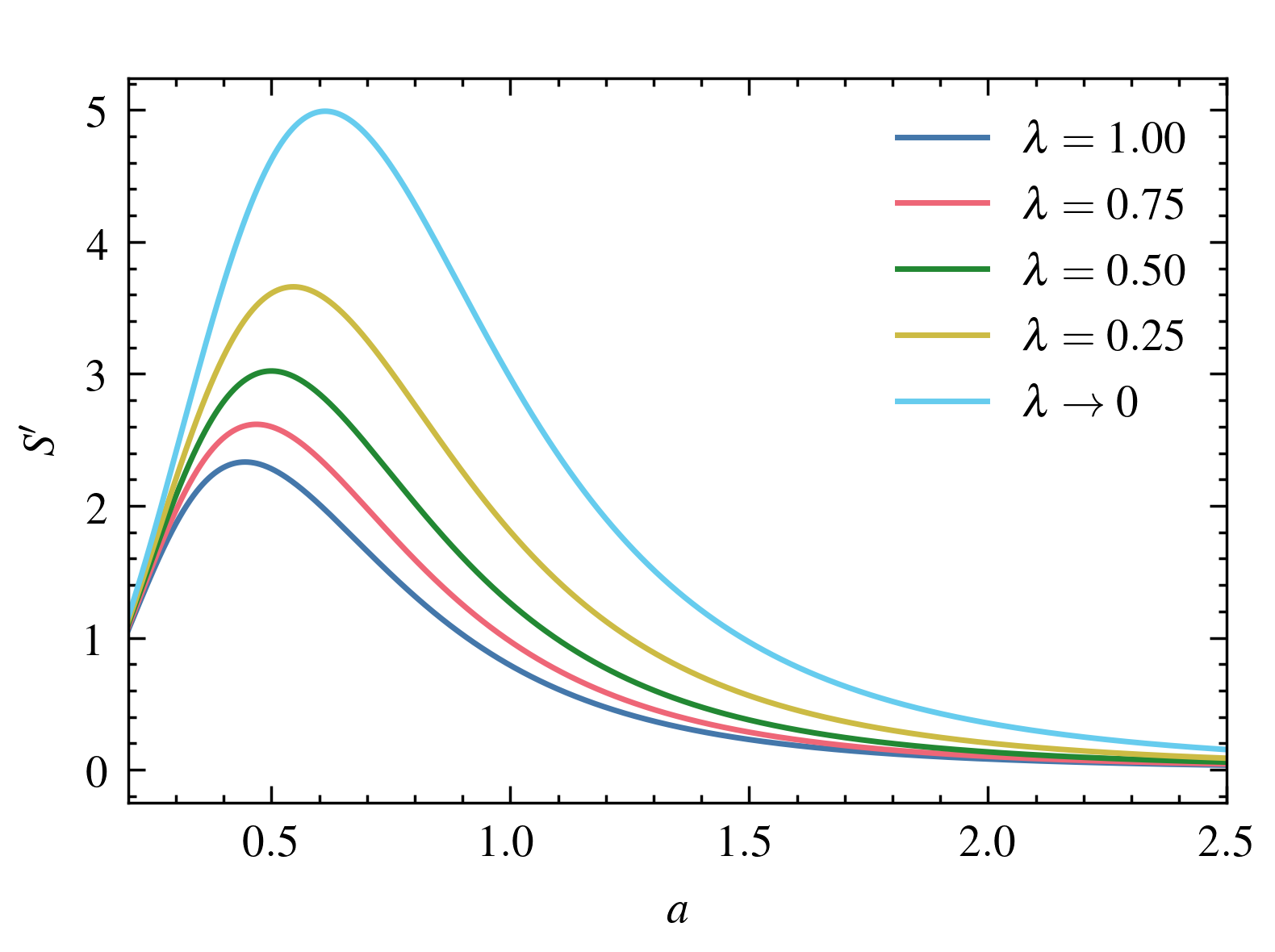}
	\includegraphics[width=0.45\textwidth]{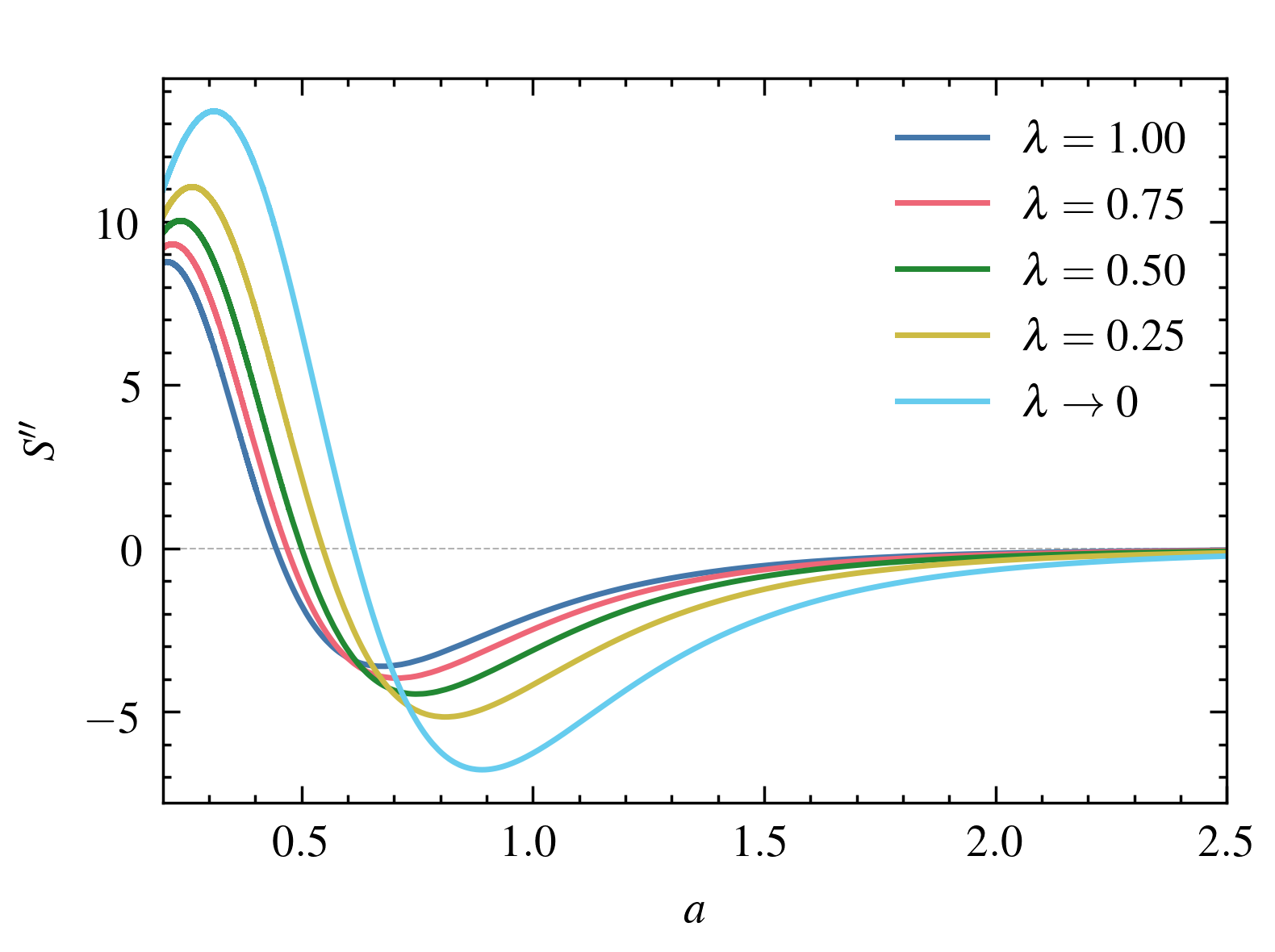}
	\caption{First ($S'$) and second ($S''$) derivatives of horizon entropy with respect to scale factor as a function of $a$ for different values of $\lambda$ (with $\Omega_{m0}=0.315$ and $H_0=67.4$ km/s/Mpc \cite{aghanim2020planck}). Here, $\lambda\rightarrow0$ implies $\Lambda$CDM behaviour (in natural units).}
	\label{fig:13}
\end{figure}

For clarity, we plot entropy as a function of the scale factor ($a$). From figure (\ref{fig:12}), we can see that the entropy tends to saturate at a maximum value as we go forward in time. To make this evident, we calculate the first and second derivatives of entropy with respect to the scale factor. The first derivative reads,
\begin{equation}
S'= \frac{-2 \pi}{H \left(H^{2} + \pi \lambda\right)}\left(\frac{dH}{da}\right).
\end{equation}
Since $dH/da$ is a negative function, as $H$ decreases with respect to time in our model, $S'$ is a positive valued function. Further, as $S$ tends to attain a maximum, the first derivative must tend to zero. These features are evident from the figure (\ref{fig:13}).
To confirm the maximization of horizon entropy, we evaluate the second derivative with respect to the scale factor, and  it reads,
\begin{align}
S''=& - \frac{2 \pi \frac{d^{2}H}{d a^{2}} }{\left(\pi \lambda + H^{2}{\left(a \right)}\right) H} + \frac{2 \pi \left(\frac{dH}{d a} \right)^{2}}{\left(\pi \lambda + H^{2}{\left(a \right)}\right) H^{2}{\left(a \right)}} \nonumber\\&+ \frac{4 \pi \left(\frac{dH}{d a} \right)^{2}}{\left(\pi \lambda + H^{2}{\left(a \right)}\right)^{2}}
\end{align}
Here, the last two terms are positive, while the first term is negative for a model where $H$ continuously decreases in the future direction. In the long run, the last two terms dilute away faster, and the first term becomes dominant. Thus, in the long run, the second derivative is negative, and we can confirm that the horizon entropy saturates at a maximum value (See figure (\ref{fig:13})). Thus our model is thermodynamically consistent and respects the second law of thermodynamics.

\section{Conclusions}

The holographic principle and the laws of thermodynamics are well-established frameworks to explain the evolution of the universe. Since the thermodynamic properties of the cosmic horizon are the fundamental basis of the holographic principle, we explored the possible connection between holographic dark energy and the laws of thermodynamics. First, we explored Moradpour et al.'s idea in \cite{Moradpour2018}, where they proposed a holographic dark energy from the principles of thermodynamics. If we were to start from the first law of horizon thermodynamics, we found that one cannot satisfy the proposed proportionality by Moradpour et al. for all choices of entropies. Their proposal appears consistent with the standard holographic approach only when the entropy follows an exponent stretched area law. One can artificially work around this problem by redefining the horizon temperature. Although such a strategy appears inconsistent with the standard laws of thermodynamics and is premature in the current stage of research, there are promising attempts via modified gravity theories \cite{nojiri2022alternative}.

Similar to the standard holographic dark energy approach, Moradpour et al.'s proposal assume the validity of the standard Friedmann equations. However, one can only derive the standard Friedmann equations from the first law of thermodynamics if we consider Bekenstein-Hawking entropy as the horizon entropy. Thus, as Golanbari~et al. point out in \cite{golanbari2020renyi}, it is crucial to consider modified Friedmann equations when using entropies other than the standard area law. In other words, from the thermodynamic perspective, it is inconsistent with using the standard Friedmann equations and the continuity relation simultaneously for entropies other than the Bekenstein-Hawking entropy. 

Now, to see if this disparity between the definition of dark energy densities by holography and thermodynamics is due to the assumption of standard Friedmann equations, we derived the expression for dark energy density from the Clausius relation and the unified first law for a dark energy dominated FLRW universe. Instead of assuming the validity of standard Friedmann equations, we took the continuity relation and the first law as the principle equations. Once again, we found it challenging to construct a one-to-one correspondence between the standard HDE and the dark energy stipulated by the first law of thermodynamics. Additionally, although the Hubble horizon is a well-established thermodynamic surface using which one can explain the cosmic evolution from the laws of thermodynamics, in the standard HDE model, the Hubble horizon cannot explain the effects of dark energy unless we invoke some phenomenological interactions. In a recent paper, R. G. Landim \cite{PhysRevD.106.043527} points out that the Hubble scale cutoff with certain phenomenological interactions fails to explain the matter and the CMB power spectrums. Thus, the standard holographic and thermodynamics approaches may not go hand in hand.

Further, the standard HDE approach with the Hubble horizon has additional issues. We first discuss the R\'enyi HDE using the conventional holographic approach to see these problems explicitly. Although the model could explain late-time accelerated expansion, it suffers from inconsistencies due to the assumption of the standard Friedmann equations and the continuity equation simultaneously with the Hubble horizon as the IR cutoff. We saw that the total equation of state parameter has different expressions when considering the validity of standard Friedmann equations and that of the continuity equation. Further, evaluating the dark energy equation of state parameter, we found that it crosses the phantom divide, which may not be a favourable behaviour \cite{PhysRevD.71.023515}, as we do not consider any interaction between the dark energy and matter.

Additionally, the model cannot explain the matter-dominated epoch when we use the continuity equation to define the equation of state parameter. A possible reason might be the non-vanishing behaviour of the dark energy density parameter. We found that for the above R\'enyi HDE model, the dark energy density parameter settles at a constant value instead of diluting to zero for large values of $z$. This behaviour effectively generates a non-zero pressure, resulting in the wrong equation of state parameter. 

On the other hand, in the simplest scenario, it is possible to induce dark energy as an integration constant using the laws of thermodynamics on the Hubble horizon. Thus, the thermodynamic approach does not demand any additional dark energy component. These observations led us to reconsider the standard HDE approach from the laws of thermodynamics.

In the thermodynamic approach, when considering entropies other than the Bekenstein-Hawking entropy, we can set this integration constant to zero by the proper definition of a dark energy component. We remove the contribution from the standard area law from the total energy density to achieve this dark energy density. This definition is consistent with the original CKN ``bound'', as we remove the contribution by the Bekenstein-Hawking term. Interestingly this definition reduces to the cosmological constant in the $\Lambda$CDM model when the generalized entropy reduces to the Bekenstein-Hawking entropy. Thus, if dark energy is dynamical, our model gives a simple extension to the $\Lambda$CDM model from the laws of thermodynamics, which is consistent with the holographic principle. 

We then demonstrate the validity of our model with R\'enyi entropy as our choice by investigating different cosmological parameters. Our model can explain the late-time acceleration and gives a final de Sitter epoch. In our model, the dark energy equation of state parameter never crosses the Phantom divide, and the dark energy density parameter reduces to zero for large $z$. This behaviour thus guarantees the proper matter-dominated and dark energy-dominated epochs. By analysing the sound speed squared, we found that the model goes to a final stable state, and the statefinder diagnostic tool can distinguish the model from the $\Lambda$CDM model. The numerical estimates of various parameters are well within the observational bounds for $\lambda\in[0,1]$. Finally, the model is consistent with the second law of thermodynamics, by which the entropies tend to a maximum towards the future.

In future, we plan to investigate the effects of phenomenological interaction terms in the continuity relation and the ability of the model to explain the observed power spectrums. It is also promising to look for solutions to problems such as the Hubble tension with our model or some extended versions of the same, as several varying dark energy models could give a reasonable explanation to it \cite{Li_2019}. Whether this demands a phantom or Quintessence behaviour is still an open area of research \cite{PhysRevD.103.L081305, Lee_2022}. Finally, it will be interesting to look for a possible connection between our approach and other holographic approaches, such as Padmanabhan's law of emergence of cosmic space \cite{padmanabhan2012emergence}, as the footings of his idea are the laws of thermodynamics. 

\begin{acknowledgements}
Authors take this opportunity to thank the referees for their valuable comments and suggestions to improve the quality of this manuscript. Thanks to Eoin O Colgain and Tanmoy Paul for pointing out useful references related to the Phantom aspect of HDE. The authors gratefully acknowledge fruitful conversations with Hassan Basari V. T. and P. B. Krishna on thermodynamic aspects of gravity. Thanks to Sarath N., Nandhida Krishnan P. and M. Dheepika for valuable discussions on dynamical dark energy models. Manosh T. Manoharan acknowledges CSIR-NET-JRF/SRF, Government of India, Grant No: 09 / 239(0558) / 2019-
EMR-I for the financial support.  
\end{acknowledgements}

\appendix

\section{R\'{e}nyi modified Friedmann equations}\label{AppendixA}

In this section, let us first construct the modified Friedmann equations from the laws of thermodynamics using R\'enyi entropy. This is to supplement the modified equations used in the previous discussion. As described earlier, for Clausius relation $-dE=TdS$, we consider the energy flux crossing the horizon rather than the total gravitating energy. Thus we have,
\begin{equation}
-dE=n\Omega_n{\tilde{r}_{\!\!_A}}^{n-1}(\rho+p)H{\tilde{r}_{\!\!_A}} dt
\end{equation}
Now applying the continuity relation in equation (\ref{eq:continutity_N}) we get,
\begin{align}
-dE&=-\Omega_n{\tilde{r}_{\!\!_A}}^nd\rho\\
\implies-\Omega_n{\tilde{r}_{\!\!_A}}^nd\rho&=TdS\implies\rho=-\int\frac{1}{\Omega_n{\tilde{r}_{\!\!_A}}^n}TdS
\end{align}
An interesting aspect about the thermodynamic approach is that, even if we consider an $n+1$ dimensional system, the temperature is still $T=1/(2\pi{\tilde{r}_{\!\!_A}})$. This is because the surface gravity depends on $h_{ab}$ rather than the full FLRW metric. All the other $n-1$ spatial dimensions contributes to the surface metric $d\Omega_{n-1}$.

Now the expression for R\'{e}nyi entropy used in black hole physics and cosmology is,
\begin{align}
S_{_\text{R}}=\frac{1}{\lambda}\log\left(1+\lambda\frac{n\Omega_{n}{\tilde{r}_{\!\!_A}}^{n - 1}}{4}\right)
\end{align}
where, $A=n\Omega_{n}{\tilde{r}_{\!\!_A}}^{n - 1}$ \cite{Cai_2005} and $\lambda$ is the extra parameter. Now we have,
\begin{equation}
TdS = \frac{n\Omega_{n}  {\tilde{r}_{\!\!_A}}^{n - 3} \left(n - 1\right)}{8 \pi \left(\frac{\Omega_{n} \lambda n {\tilde{r}_{\!\!_A}}^{n - 1}}{4} + 1\right)}d{\tilde{r}_{\!\!_A}}.
\end{equation}
Now using the Clausius relation, we get,
\begin{align}
&2\Omega_n\rho=-\int\frac{n\Omega_{n}  {\tilde{r}_{\!\!_A}}^{- 3} \left(n - 1\right)}{\pi \left({\Omega_{n} \lambda n {\tilde{r}_{\!\!_A}}^{n - 1}} + 4\right)}d{\tilde{r}_{\!\!_A}}
\nonumber\\&\implies 2\Omega_{n}\rho=\frac{1-n}{\pi\lambda}\int\left(\frac{{\tilde{r}_{\!\!_A}}^{-3}}{{\tilde{r}_{\!\!_A}}^{n-1}+\zeta}\right)d{\tilde{r}_{\!\!_A}}
\end{align}
Here, we defined, $\zeta={4}({n\Omega_{n}\lambda})^{-1}$. On integrating the above equation, we will get the first modified Friedmann equation as,
\begin{align}
2\Omega_{n}\rho=\frac{n-1}{\pi\lambda(n+1)}\frac{1}{{\tilde{r}_{\!\!_A}}^{(n+1)}} {}_2{\rm F}_1\left(1,\frac{n+1}{n-1};\frac{2n}{n-1};-{\tilde{r}_{\!\!_A}}^{1-n}\zeta\right)\label{eq:genRFE1}
\end{align}
Here, ${}_2{\rm F}_1\left(1,\frac{n+1}{n-1};\frac{2n}{n-1};-{\tilde{r}_{\!\!_A}}^{1-n}\zeta\right)$ is the hyper-geometric function \cite{olver2010nist}. To obtain the second Friedmann equation, one can differentiate the first Friedmann equation, use the generalised continuity equation, and then substitute for $\rho$ using the first modified Friedmann equation. Thus we have,
\begin{align}
&-2\Omega_{n}p\nonumber\\&=\left(\frac{n-1}{\pi\lambda(n+1)}\frac{1}{{\tilde{r}_{\!\!_A}}^{(n+1)}} {}_2{\rm F}_1\left(1,\frac{n+1}{n-1};\frac{2n}{n-1};-{\tilde{r}_{\!\!_A}}^{1-n}\zeta\right)+\right.\nonumber\\&\left.\frac{n-1}{\pi\lambda\left(n+1\right)nH}\frac{d}{dt}\left\lbrace\frac{1}{{\tilde{r}_{\!\!_A}}^{(n+1)}} {}_2{\rm F}_1\left(1,\frac{n+1}{n-1};\frac{2n}{n-1};-{\tilde{r}_{\!\!_A}}^{1-n}\zeta\right)\right\rbrace\right)\label{eq:genRFE2}
\end{align}

Equations (\ref{eq:genRFE1}) and (\ref{eq:genRFE2}) are the general $n+1$ non-flat R\'{e}nyi modified Friedmann equations. When $n=3$ we get,
\begin{align}
\frac{8\pi}{3}\rho=\displaystyle H^{2}+ \frac{k}{a^{2}} + \pi \lambda \log{\left(1+\frac{{H^{2} + \frac{k}{a^{2}}}}{\pi\lambda} \right)} 
\end{align}
\begin{align}
-\frac{8\pi}{3}p=&~\displaystyle H^{2}+ \frac{k}{a^{2}} + \pi \lambda \log{\left(1+\frac{{H^{2} + \frac{k}{a^{2}}}}{\pi\lambda} \right)} \nonumber\\&+ \frac{2 \left(H^{2} \dot{H} a^{4} - H^{2} a^{2} k + \dot{H} a^{2} k - k^{2}\right)}{3 a^{2} \left(H^{2} a^{2} + \pi \lambda a^{2} + k\right)}
\end{align}

Considering the recent observational constraints \cite{brout2022pantheon+,aghanim2020planck}, the curvature $k\approx0$. Thus it is reasonable to consider flat FLRW spacetime. For the simplest case of $3+1$ flat FLRW spacetime, the R\'{e}nyi modified Friedmann equations becomes,
\begin{equation}
3H^2+3\lambda\pi\log\left(1+\frac{H^2}{\pi\lambda}\right)=8\pi  \rho\label{eq:RFE11}
\end{equation}
\begin{equation}
\frac{2 H^{2} \dot{H}}{\left(H^{2} + \pi \lambda\right)}+3H^2+3\lambda\pi\log\left(1+\frac{H^2}{\pi\lambda}\right)=-8\pi p
\label{eq:RFE21}
\end{equation}
Now using the above expressions, we can discuss various cosmological models. One may consider this the governing equation and add the dark energy component to $\rho$. Alternatively, as pointed out in Ref. \cite{PhysRevD.105.044042, NOJIRI2022137189}, one may rewrite this in the form of a standard Friedmann equation and redefine pressure and density.

\end{document}